\documentclass[pdflatex,sn-aps, iicol]{sn-jnl}

\usepackage{graphicx}%
\usepackage{multirow}%
\usepackage{amsmath,amssymb,amsfonts}%
\usepackage{amsthm}%
\usepackage{mathrsfs}%
\usepackage[title]{appendix}%
\usepackage{xcolor}%
\usepackage{textcomp}%
\usepackage{manyfoot}%
\usepackage{booktabs}%
\usepackage{algorithm}%
\usepackage{algorithmicx}%
\usepackage{algpseudocode}%
\usepackage{listings}%
\usepackage{graphicx}
\usepackage{subcaption}

\theoremstyle{thmstyleone}%
\theoremstyle{thmstyletwo}%

\theoremstyle{thmstylethree}%

\begin{document}

\title[Anisotropy of the Universe]{Revisiting cosmic anisotropy with the Pantheon+ compilation}

\author[1]{\fnm{Yong} \sur{Zhou}}

\author*[2]{\fnm{Dong} \sur{Zhao}}\email{zhaodong@yzu.edu.cn}

\affil[1]{Department of Applied Physics, College of Science, China Agricultural University, 17 Qinghua East Road, Haidian District, Beijing 100083, China}

\affil[2]{Center for Gravitation and Cosmology, College of Physical Science and Technology, Yangzhou University, Yangzhou, 225009, China}

\abstract{

We investigate cosmic anisotropy within the updated Pantheon+ sample using both the dipole fitting (DF) and hemisphere comparison (HC) methods. With the DF method, the dipole signal within the full sample is statistically weak. However, the low-$z$ subsample yields a dipole signal of $A_{\mathrm{D}} = 0.952^{+0.454}_{-0.403} \times 10^{-3}$ at $\sim 2\sigma$ significance, pointing towards $(l,b) = (149.77^\circ, -12.20^\circ)$. This signal is predominantly driven by a combined subset of surveys 5, 56, 63, and 150, which is characterized by an amplitude of $A_{\mathrm{D}} = 1.730_{-0.715}^{+0.554} \times 10^{-3}$ towards $(l,b) = (153.05^\circ, -1.25^\circ)$. For the HC method, the full sample yields a maximum anisotropy level of $\mathrm{AL}_{\mathrm{max}} = 0.289 \pm 0.052$ oriented towards $(l,b) = (127.97^\circ, 17.90^\circ)$ with a $1.56\sigma$ significance. This preferred direction is primarily determined by the highly inhomogeneous SNLS subsample, whereas the low-$z$ and high-$z$ subsamples act to suppress the anisotropy level along this axis. These subsample-dependent results suggest that the apparent anisotropy arises from local structures or the inhomogeneous distribution of the datasets rather than an intrinsic cosmic anisotropy.

}


\maketitle

\section{Introduction}\label{sec1}
Modern cosmology relies on the cosmological principle, which posits large-scale homogeneity and isotropy across the Universe. Precise measurements of the cosmic microwave background (CMB) by the WMAP \citep{WMAP:2012nax} and Planck \citep{Planck:2018vyg} satellites have strongly supported this assumption. Nonetheless, various observational anomalies have emerged, potentially challenging this fundamental hypothesis. Notable examples include the hemispherical power asymmetry \citep{Eriksen:2003db, Hansen:2004vq, WMAP:2012fli, Akrami:2014eta, Quartin:2014yaa, Planck:2019evm} and parity violation \citep{Kim:2010gf, Kim:2010gd, Gruppuso:2010nd, Kim:2010st, Zhao:2013jya} in the CMB, and the alignment of quasar polarization vectors over large scales \citep{Hutsemekers:2005iz, Pelgrims:2016zbr}. Such unexpected features hint at the existence of a preferred direction in the cosmic expansion.

As standard candles, Type Ia supernovae (SNe Ia) provide a powerful tool to probe cosmic anisotropy. A $2\sigma$ dipole signal was initially detected in the Union2 and Union2.1 datasets in the Union2 and Union2.1 datasets \citep{Mariano:2012wx, Zhao:2013yaa, Yang:2013gea, Li:2015uda, Bengaly:2015dza, Javanmardi:2015sfa, Lin:2016jqp}. However, this signal vanished in the larger JLA \citep{Bengaly:2015dza, Lin:2015rza, Wang:2017ezt, Chang:2017bbi, Deng:2018yhb, Sun:2018epo, Rahman:2021mti} and Pantheon compilations \citep{Sun:2018cha, Deng:2018jrp, Andrade:2018eta, Li:2019bqb, Zhao:2019azy, Chang:2019utc}. Nonetheless, using the hemisphere comparison (HC) method, a spatial variation in the matter density was detected within the Pantheon sample at the $2.1\sigma$ level \citep{Zhao:2019azy}. Several studies suggest that these anisotropic signals may be influenced by the non-uniform sky coverage of the datasets. For instance, the anisotropic signals detected within the Pantheon sample has been shown to be highly sensitive to the inhomogeneous spatial distribution of SNe Ia \citep{Zhao:2019azy}. Similarly, coordinate anisotropy has been shown to shift the dipole direction and increase its magnitude \citep{Sun:2018epo}.

Currently, the Pantheon+ compilation \citep{Scolnic:2021amr} stands as a comprehensive sample of SNe Ia. Using the dipole fitting (DF) method, the full dataset remains consistent with large-scale isotropy \citep{Tang:2023kzs, Bengaly:2024ree, Yang:2024rsy}. However, a dipole signal with a statistical significance of approximately $2\sigma$ emerges in the low-redshift regime \citep{Tang:2023kzs, Sorrenti:2022zat, Cowell:2022ehf, Sorrenti:2024ugq, Sah:2024csa}. Conversely, the HC method reveals notable anisotropic signals within the Pantheon+ sample \citep{Hu:2023eyf, Hu:2024qnx}, with a sharp change in the anisotropy level occurring at distances below 40 Mpc \citep{Perivolaropoulos:2023tdt}.

In our previous work, we investigated the contribution of individual subsamples to cosmic anisotropy within the Pantheon compilation. The DF analysis indicated that the SDSS subsample plays a key role in the dipole anisotropy of the full sample. Meanwhile, the HC analysis showed that the overall anisotropy is primarily driven by the Low-$z$ and SNLS subsamples \citep{Zhao:2019azy}. In this study, we analyze the influence of individual subsamples on cosmic anisotropy using the updated Pantheon+ dataset. The remainder of this paper is structured as follows. Section \ref{sample} introduces the Pantheon+ sample. We then employ the DF and HC methods to analyze cosmic anisotropy in Sections \ref{DF} and \ref{HC}, respectively. Finally, our main conclusions are summarized in Section \ref{Conclusion}.

\section{Pantheon+ sample}\label{sample}
\begin{figure}
	\begin{center}
		\includegraphics[width=7.5cm]{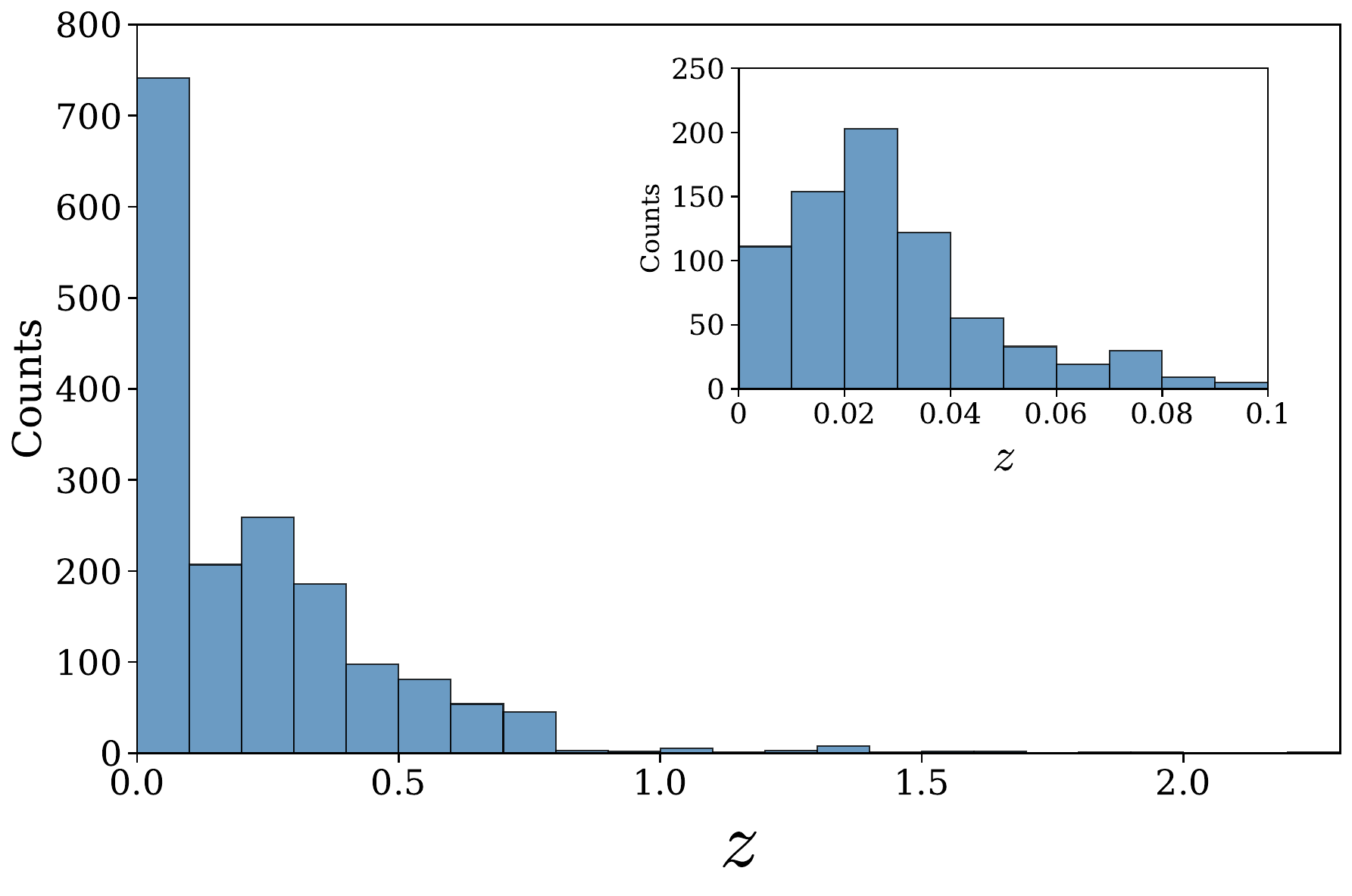}
		\caption{Redshift ($z_{\mathrm{hd}}$) distribution of the Pantheon+ sample, with the inset highlighting the low-redshift SNe Ia at $z < 0.1$.}
		\label{fig:red_all}
	\end{center}
\end{figure}

\begin{figure}
	\begin{center}
		\includegraphics[width=7.5cm]{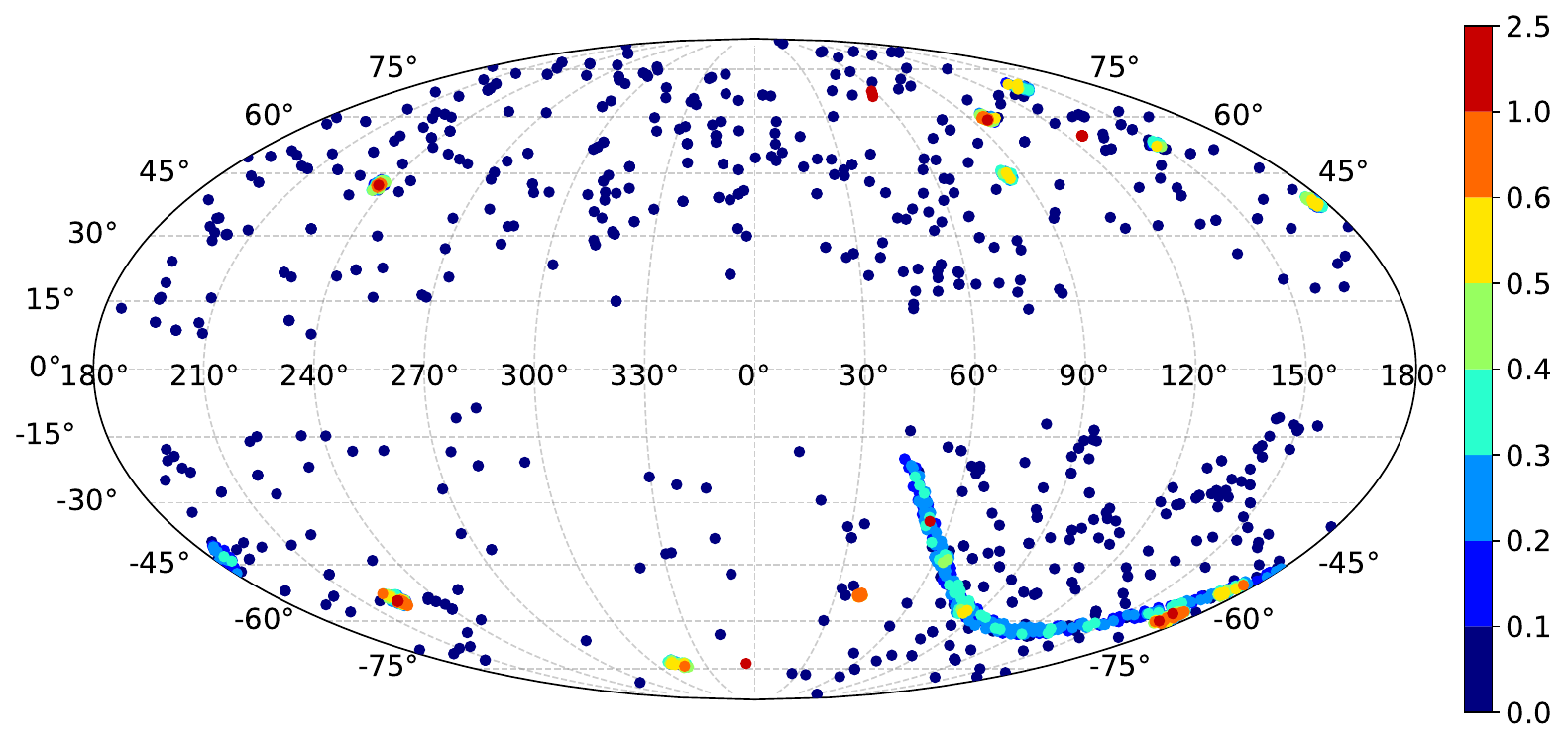}
		\caption{Distributions of the Pantheon+ sample in Galactic coordinates, with the color scale representing the redshift ($z_{\mathrm{hd}}$) of the SNe Ia.}
		\label{fig:lb_all}
	\end{center}
\end{figure}

Scolnic et al. \citep{Scolnic:2021amr} presented the Pantheon+ compilation, consisting of 1701 light curves from 1550 unique SNe Ia spanning the redshift range of $0.001 < z < 2.27$. Compared to its predecessor, the Pantheon+ dataset contains a substantially larger number of low-redshift SNe Ia. Figure \ref{fig:red_all} displays the redshift distribution of the Pantheon+ sample, while Figure \ref{fig:lb_all} illustrates its spatial distribution in Galactic coordinates.

\begin{figure}
	\begin{center}
		\includegraphics[width=7.5cm]{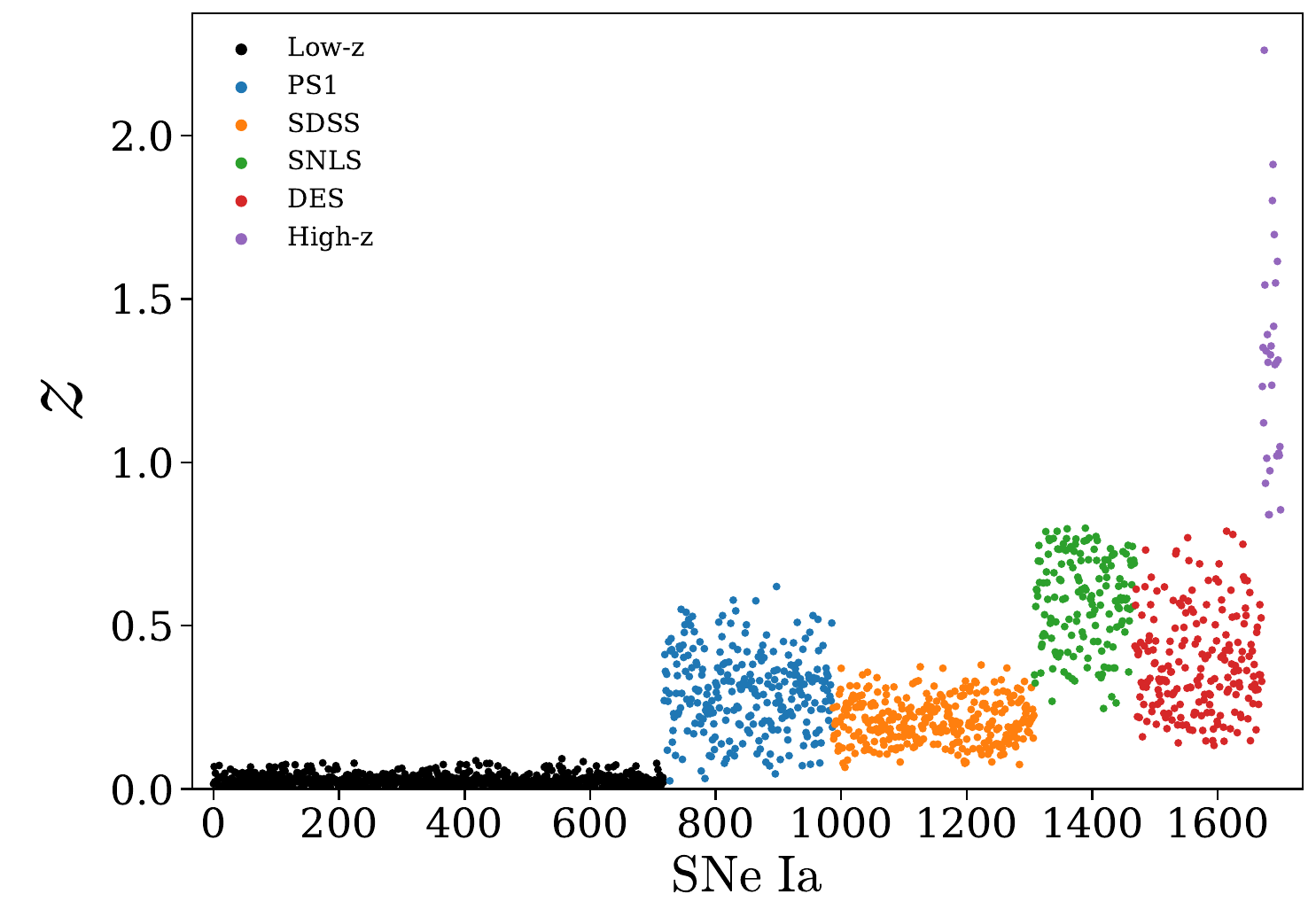}
		\caption{Redshift ($z_{\mathrm{hd}}$) distributions of the six subsamples: Low-$z$, PS1, SDSS, SNLS, DES, and High-$z$. The vertical axis represents the redshift.}
		\label{fig:red_subsample}
	\end{center}
\end{figure}

\begin{figure*}[!htbp]
	\centering
	\begin{subfigure}[b]{0.49\textwidth}
		\centering
		\includegraphics[width=\textwidth]{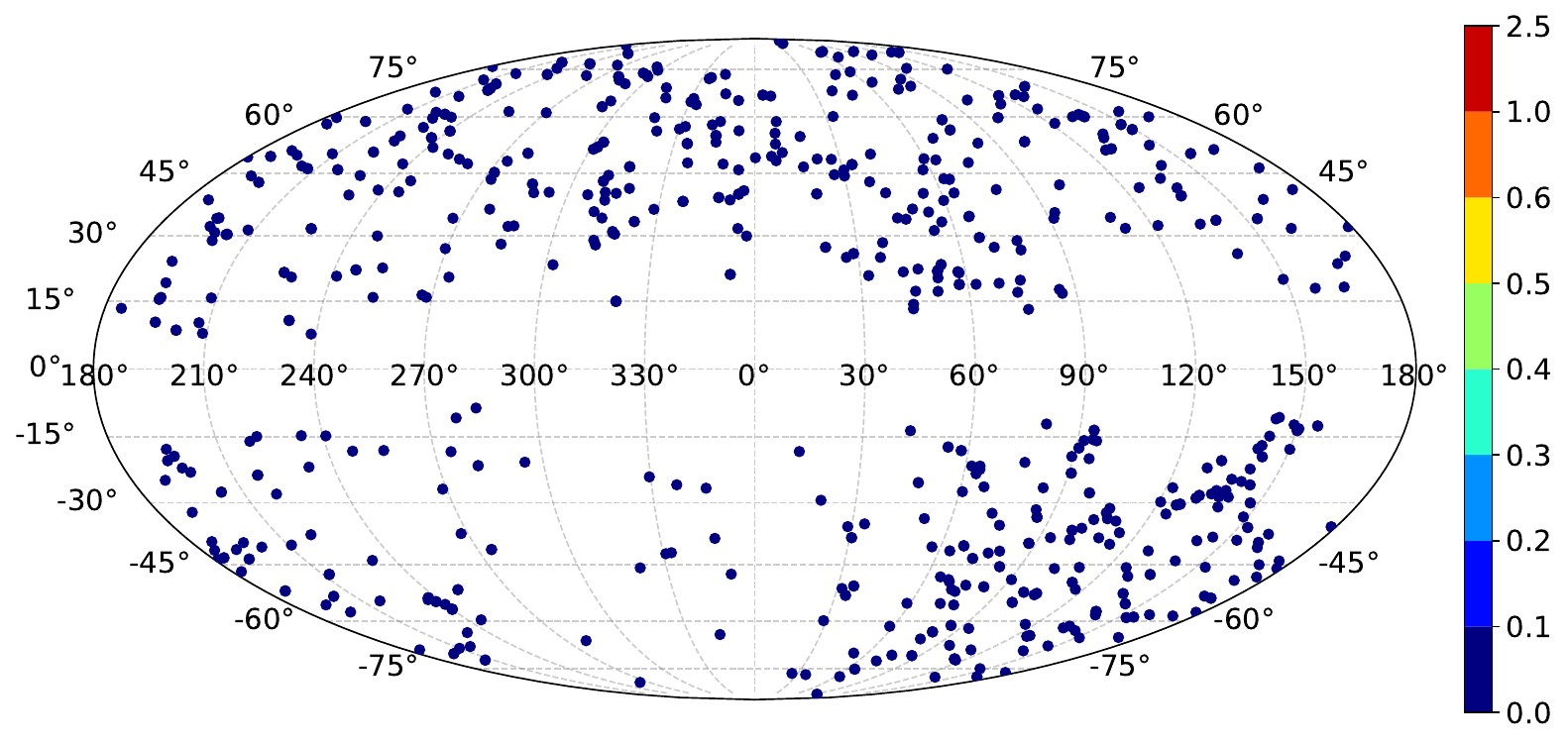}
		\caption{Low-$z$}
		\label{fig:lb_lowz}
	\end{subfigure}
	\hfill
	\begin{subfigure}[b]{0.49\textwidth}
		\centering
		\includegraphics[width=\textwidth]{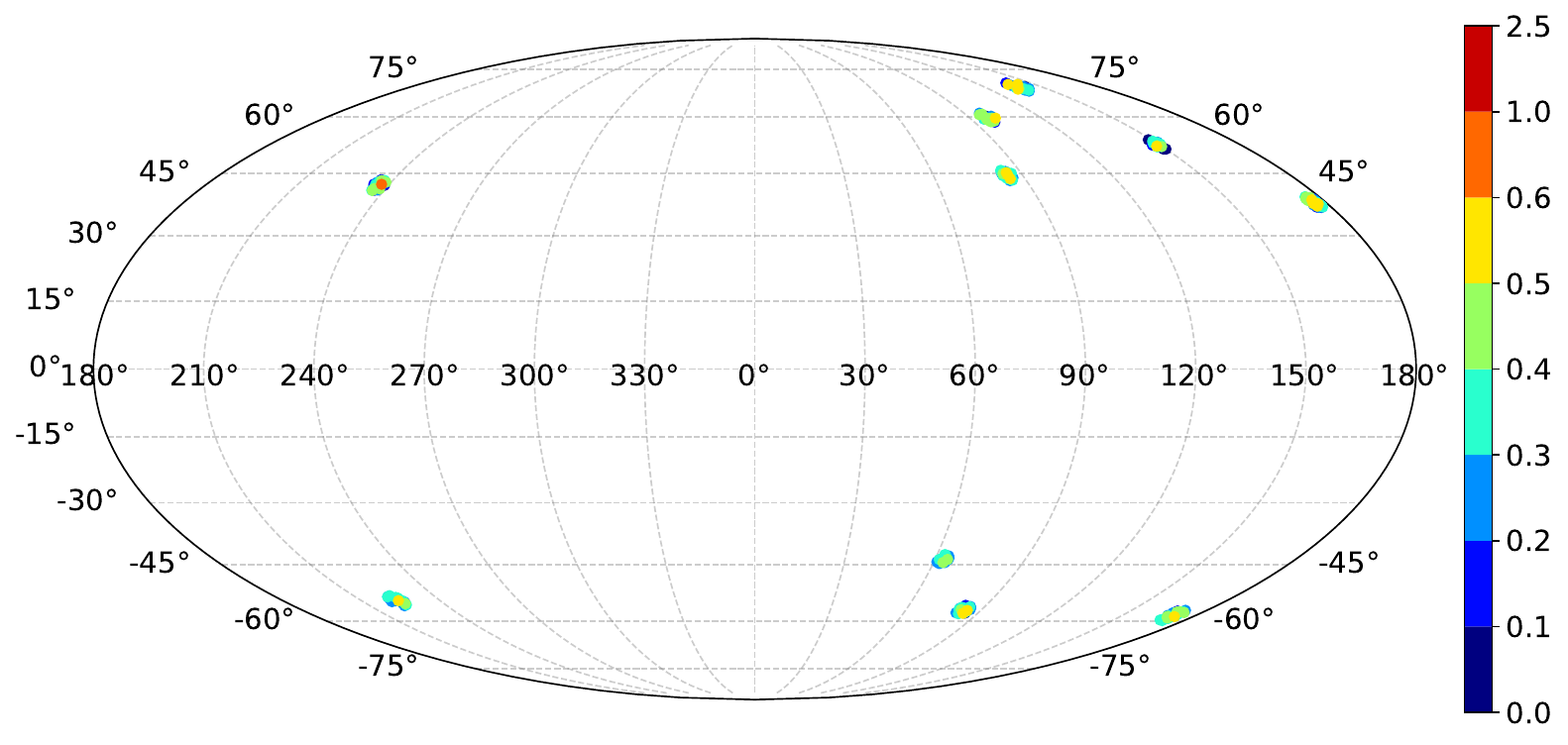}
		\caption{PS1}
		\label{fig:lb_PS1}
	\end{subfigure}
    	\begin{subfigure}[b]{0.49\textwidth}
		\centering
		\includegraphics[width=\textwidth]{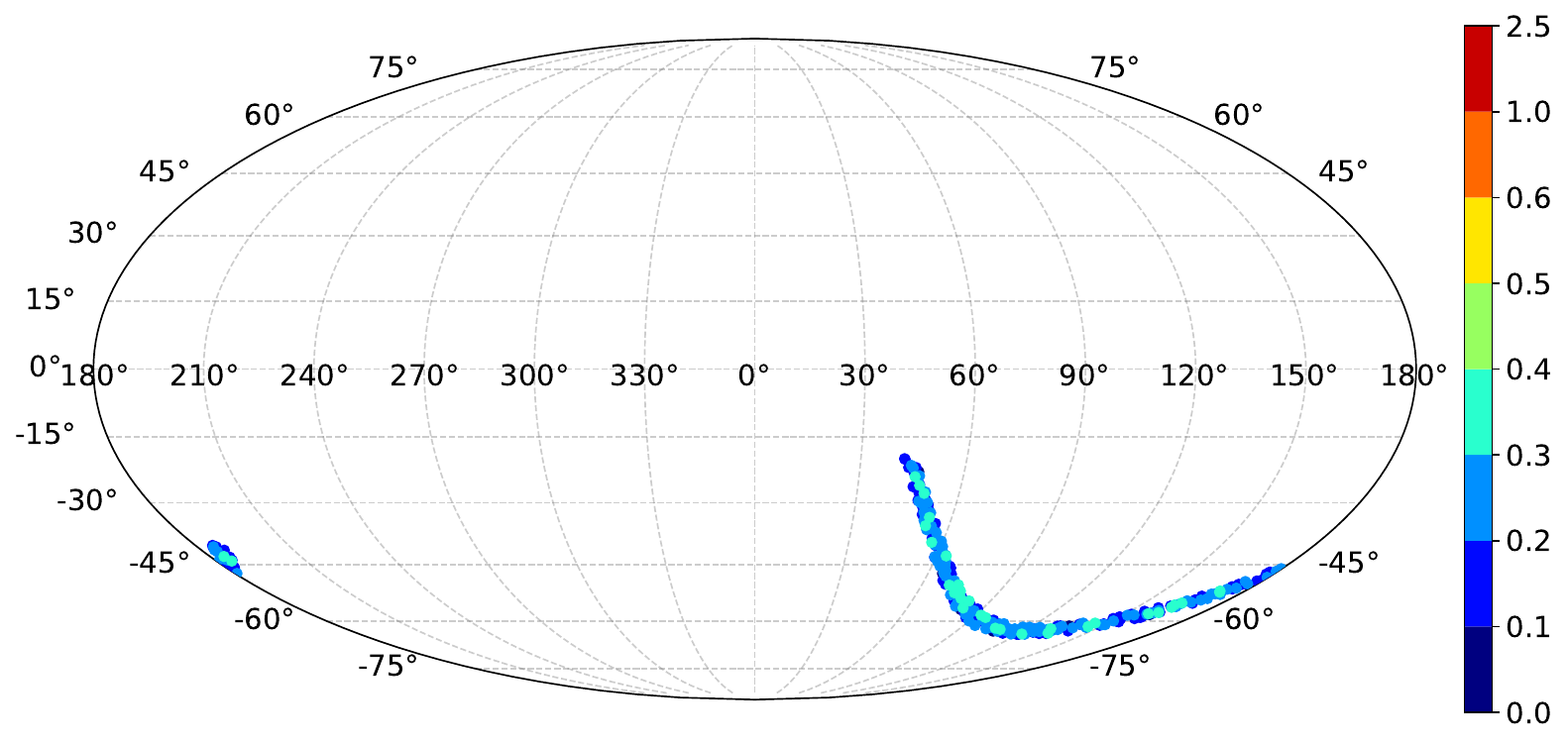}
		\caption{SDSS}
		\label{fig:lb_SDSS}
	\end{subfigure}
	\hfill
	\begin{subfigure}[b]{0.49\textwidth}
		\centering
		\includegraphics[width=\textwidth]{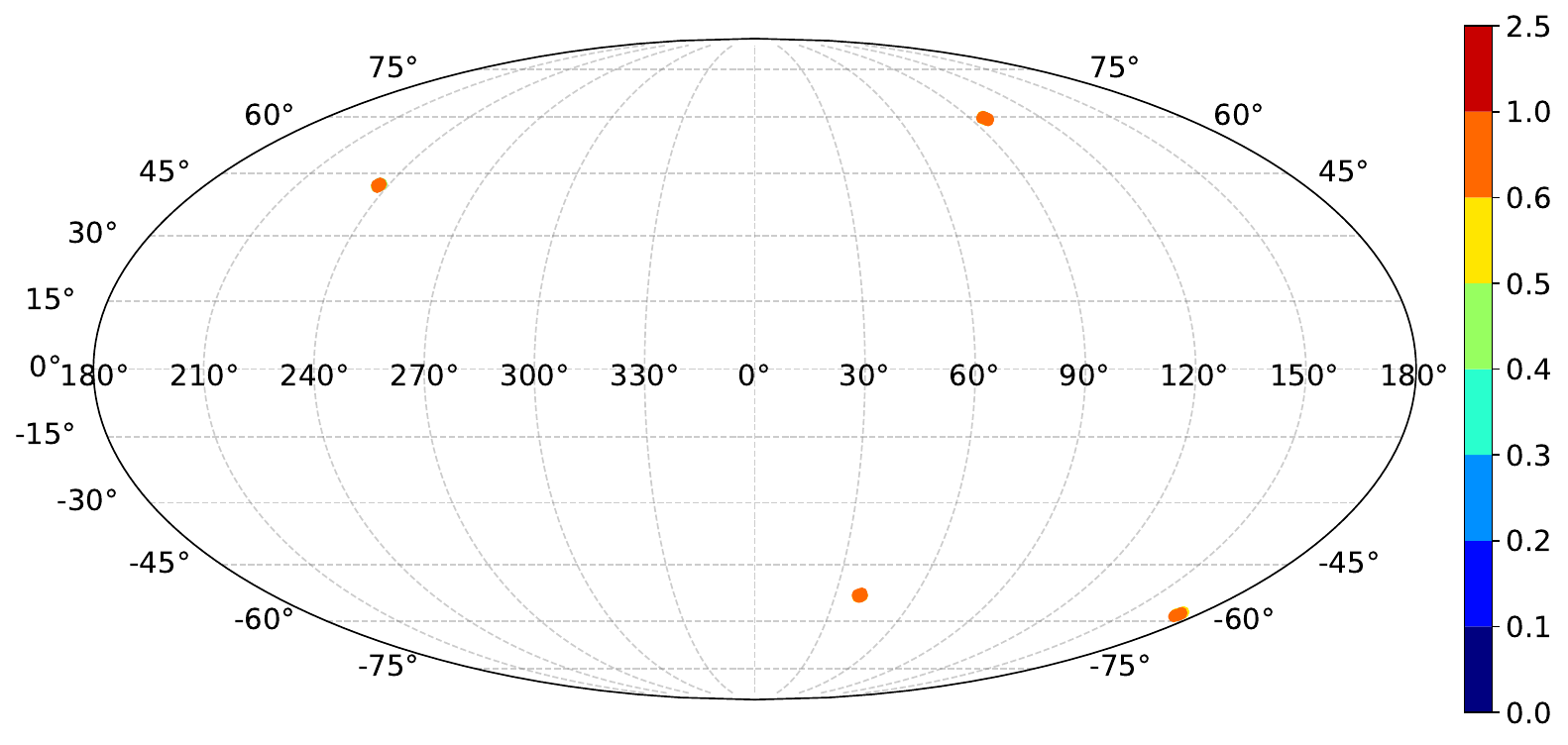}
		\caption{SNLS}
		\label{fig:lb_SNLS}
	\end{subfigure}
    	\begin{subfigure}[b]{0.49\textwidth}
		\centering
		\includegraphics[width=\textwidth]{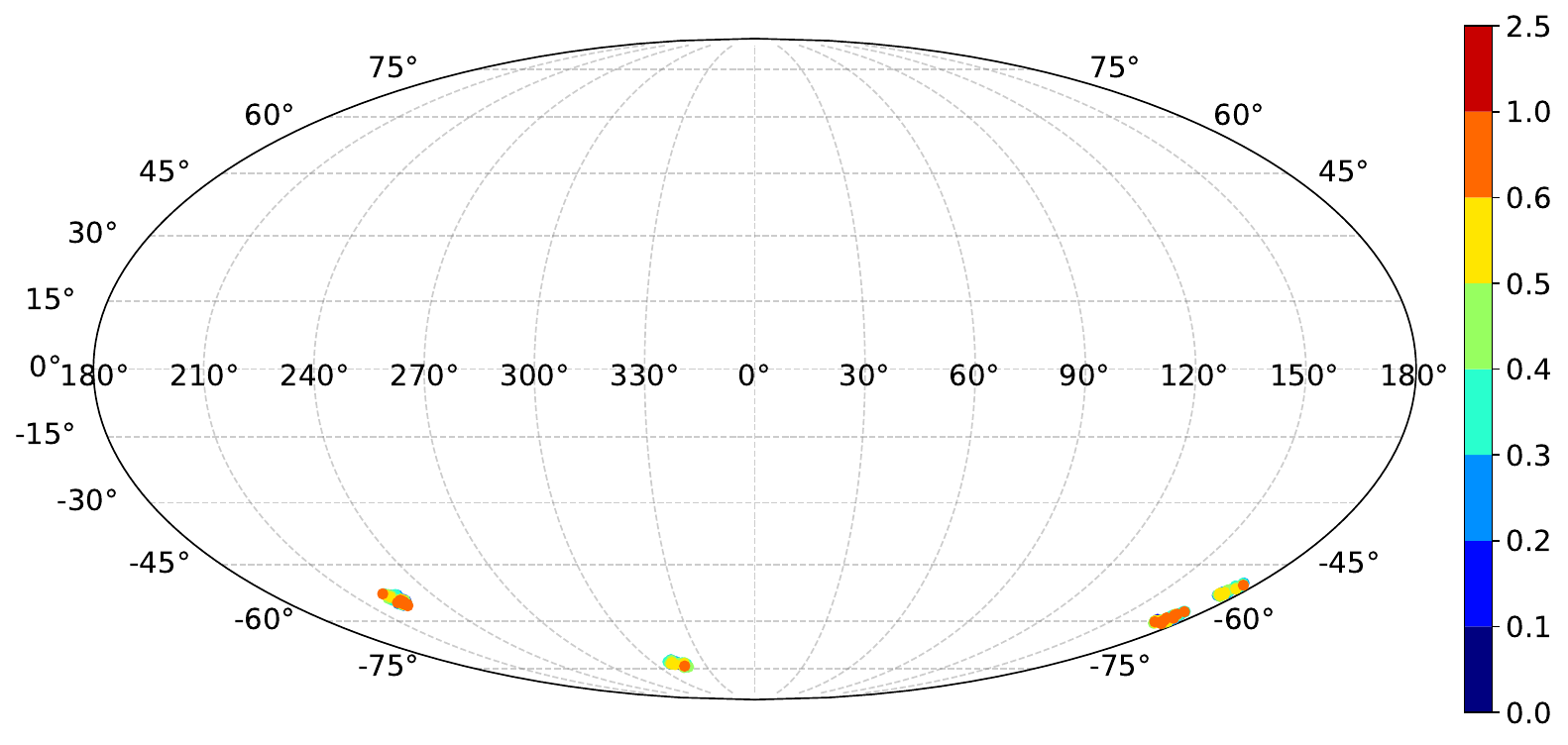}
		\caption{DES}
		\label{fig:lb_DES}
	\end{subfigure}
	\hfill
	\begin{subfigure}[b]{0.49\textwidth}
		\centering
		\includegraphics[width=\textwidth]{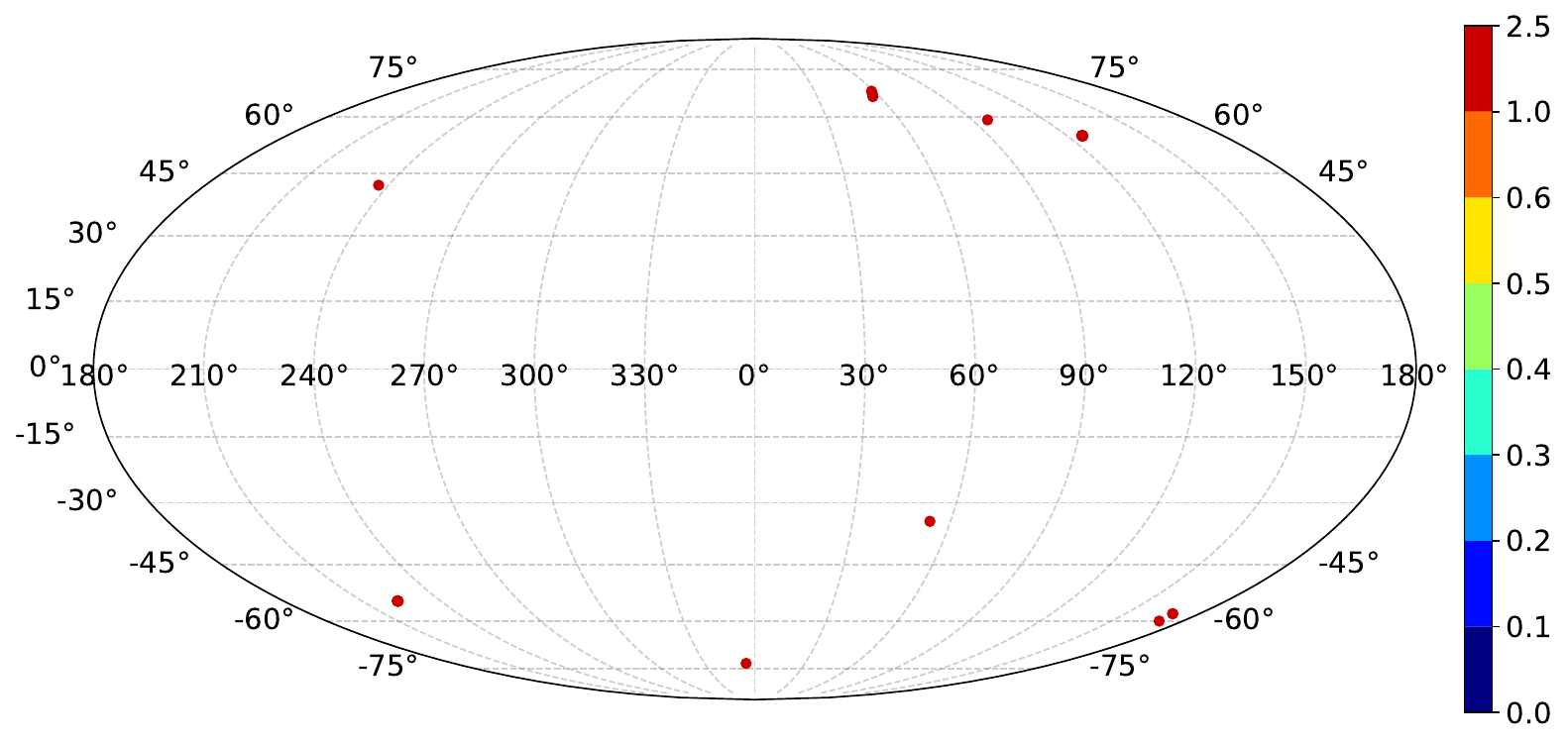}
		\caption{High-$z$}
		\label{fig:lb_Highz}
	\end{subfigure}
	\caption{Distributions of the six subsamples in Galactic coordinates, with the color scale representing the redshift ($z_{\mathrm{hd}}$) of the SNe Ia.}
	\label{fig:lb_subsample}
\end{figure*}

The Pantheon+ sample is derived from 20 distinct surveys, which we categorize into six subsamples: Low-$z$, PS1, SDSS, SNLS, DES, and High-$z$. The properties of these six subsamples, including their corresponding survey IDs, the number of SNe Ia, and the redshift ($z_{\mathrm{hd}}$) range, are summarized in Table \ref{table_subsample}. Figure \ref{fig:red_subsample} displays the redshift distributions of these subsamples, while Figure \ref{fig:lb_subsample} illustrates their spatial distributions in Galactic coordinates. The spatial distributions of these subsamples vary significantly. While the Low-$z$ subsample (718 SNe Ia) exhibits a nearly uniform sky distribution, the remaining subsamples are highly localized. Specifically, the PS1 subsample (269 SNe Ia) is distributed across approximately ten sky regions, and the DES subsample (203 SNe Ia) is concentrated within four distinct fields. Meanwhile, the High-$z$ subsample (30 SNe Ia) is scattered across nine different directions. The SDSS and SNLS subsamples exhibit highly inhomogeneous distributions. In particular, the SDSS subsample (321 SNe Ia) is confined to a narrow stripe, whereas the SNLS subsample (160 SNe Ia) is approximately concentrated across four distinct directions. Consequently, the Pantheon+ sample inherits the inhomogeneous distribution of its predecessor, although this inhomogeneity is partially mitigated by the expanded low-redshift subsample.

\begin{table*}
	\large
	\centering
	\caption{Summary of the six Pantheon+ subsamples analyzed in this study, listing the subsample name, corresponding survey IDs, the number of SNe Ia, and the redshift ($z_{\mathrm{hd}}$) range.}
	\label{table_subsample}
	\renewcommand{\arraystretch}{1.5}
	\setlength{\tabcolsep}{2.0mm}
	\resizebox{\textwidth}{!}{
		\begin{tabular}{cccc}
			\hline 
			Subsample & Survey ID & Number of SNe Ia & Redshift range \\
			\hline
			Low-$z$ 
            & $5, 18, 50, 51, 56, 57, 61, 62, 63, 64, 65, 66, 150$
            & $718$
            & $0.001 - 0.093$ \\
			PS1
            & $15$
            & $269$
            & $0.025 - 0.62$ \\
			SDSS
            & $1$
            & $321$
            & $0.066 - 0.38$ \\
			SNLS
            & $4$
            & $160$
            & $0.246 - 0.799$ \\
			DES
            & $10$
            & $203$
            & $0.133 - 0.79$ \\
			High-$z$
            & $100, 101, 106$
            & $30$
            & $0.839 - 2.27$ \\
			\hline
		\end{tabular}
	}
\end{table*}


In this study, we employ the DF and HC methods to investigate cosmic anisotropy using the Pantheon+ dataset. To evaluate the influence of individual subsamples on the overall anisotropy, we systematically exclude each subsample from the full dataset in turn. During this exclusion process, SNe Ia residing in Cepheid hosts are always retained to break the degeneracy between the Hubble constant and the absolute magnitude. The full Pantheon+ sample and its six combinations are summarized as follows:

(1) Full Pantheon+

(2) Pantheon+ excluding Low-$z$

(3) Pantheon+ excluding PS1

(4) Pantheon+ excluding SDSS

(5) Pantheon+ excluding SNLS

(6) Pantheon+ excluding DES

(7) Pantheon+ excluding High-$z$

The observed distance moduli of standardized SNe Ia can be expressed in a modified Tripp form \citep{Tripp:1997wt} as:
\begin{equation}
	\mu_{\mathrm{obs}} = m_B - M + \alpha x_1 - \beta c - \delta_{\mathrm{bias}} + \delta_{\mathrm{host}}.
\end{equation}
Here, $m_B$ represents the rest-frame $B$-band peak apparent magnitude, and $M$ is the standardized absolute magnitude of SNe Ia. The parameters $x_1$ and $c$ denote the light-curve stretch and color, respectively, while $\alpha$ and $\beta$ characterize the luminosity--stretch and luminosity--color relations. Additionally, the term $\delta_{\mathrm{bias}}$ denotes the simulation-based bias correction, whereas $\delta_{\mathrm{host}}$ accounts for the host-galaxy-mass correction.

For the Pantheon+ sample, the publicly released corrected magnitudes already incorporate the light-curve standardization corrections. The bias corrections, estimated via the BEAMS with Bias Corrections (BBC) method \citep{Kessler:2016uwi}, are also included. The associated uncertainties, including both statistical and systematic contributions, are provided through the covariance matrix. Consequently, in the cosmological analysis, one may equivalently use:
\begin{equation}
\mu_{\mathrm{obs}}=m_{\mathrm{obs}}-M.
\end{equation}
The details of the Pantheon+ light-curve sample are presented in Scolnic et al. \citep{Scolnic:2021amr}, while the corresponding distance measurements, bias corrections, and cosmological likelihood are described in Brout et al. \citep{Brout:2022vxf}.

In the standard cosmological model, the theoretical distance modulus is given by:
\begin{equation}\label{modu_th}
\mu_{\mathrm{th}}=5 \log \frac{d_{\mathrm{L}}}{\mathrm{Mpc}}+25,
\end{equation}
where $d_{\mathrm{L}}$ is the luminosity distance, which is defined as:
\begin{equation}\label{dL}
	d_\mathrm{L}=\frac{c(1+z_{\mathrm{hel}})}{H_0} \int_{0}^{z_{\mathrm{hd}}} \frac{d z^{\prime}}{E\left(z^{\prime}\right)},
\end{equation}
where $z_{\mathrm{hel}}$ is the heliocentric redshift, and $z_{\mathrm{hd}}$ is the Hubble diagram redshift corrected to the CMB frame and for peculiar velocities. $H_0$ represents the Hubble constant. While $M$ and $H_0$ are degenerate using SNe Ia alone, the Pantheon+ sample includes low-redshift calibrators in Cepheid hosts. Integrating SH0ES Cepheid data with SNe Ia breaks this degeneracy. In the flat $\Lambda\mathrm{CDM}$ model, $E(z)$ is given by:
\begin{equation}
	E(z)=\sqrt{\Omega_\mathrm{m}(1+z)^{3}+\left(1-\Omega_\mathrm{m}\right)},
\end{equation}
where $\Omega_\mathrm{m}$ is the matter density.

The best-fit cosmological parameters are determined by minimizing the $\chi^2$ statistic,
\begin{equation}
\chi^2 = \Delta\boldsymbol{\mu}^{T} \boldsymbol{C}^{-1} \Delta\boldsymbol{\mu},
\end{equation}
where $\boldsymbol{C}$ represents the combined statistical and systematic covariance matrix. When the SH0ES Cepheid host distances are included, the SNe Ia distance residuals $\Delta\boldsymbol{\mu}$ are modified as follows:
\begin{equation}
\Delta\boldsymbol{\mu}_i =
\begin{cases}
\mu_{{\mathrm{obs}}, i} - \mu_i^{\mathrm{Cepheid}}, & i \in {\mathrm{Cepheid\ hosts}}, \\
\mu_{{\mathrm{obs}}, i} - \mu_{\mathrm{th}}(z_i), & {\mathrm{otherwise}},
\end{cases}
\end{equation}
where $\mu^{\mathrm{Cepheid}}$ is the SH0ES Cepheid-calibrated distance to the host galaxy.

\section{Dipole fitting method and result}\label{DF}
In the dipole fitting method, the theoretical distance modulus modified by a dipole correction is expressed as:
\begin{equation}\label{dm_dipole}
	\tilde{\mu}_\mathrm{th}=\mu_\mathrm{th} \times\left(1+A_\mathrm{D}(\hat{\boldsymbol{n}} \cdot \hat{\boldsymbol{p}})\right),
\end{equation}
where $A_\mathrm{D}$ represents the dipole amplitude, while $\hat{\boldsymbol{n}}$ and $\hat{\boldsymbol{p}}$ are unit vectors denoting the dipole direction and the direction toward the source, respectively. In Galactic coordinates, the dipole direction $\hat{\boldsymbol{n}}$ is parameterized as:
\begin{equation}
	\hat{\boldsymbol{n}}=\cos (b) \cos (l) \hat{\boldsymbol{i}}+\cos (b) \sin (l) \hat{\boldsymbol{j}}+\sin (b) \hat{\boldsymbol{k}},
\end{equation}
where $l$ and $b$ correspond to the Galactic longitude and latitude of the dipole, respectively, and $\hat{\boldsymbol{i}}$, $\hat{\boldsymbol{j}}$, and $\hat{\boldsymbol{k}}$ represent the standard Cartesian unit vectors. Similarly, the unit vector $\hat{\boldsymbol{p}}_{i}$ directed toward the $i$-th source is defined as:
\begin{equation}
	\hat{\boldsymbol{p}}_{i}=\cos \left(b_{i}\right) \cos \left(l_{i}\right) \hat{\boldsymbol{i}}+\cos \left(b_{i}\right) \sin \left(l_{i}\right) \hat{\boldsymbol{j}}+\sin \left(b_{i}\right) \hat{\boldsymbol{k}}.
\end{equation}

\begin{figure*}[!htbp]
	\centering
	\begin{subfigure}[b]{0.49\textwidth}
		\centering
		\includegraphics[width=\textwidth,height=0.21\textheight,keepaspectratio]{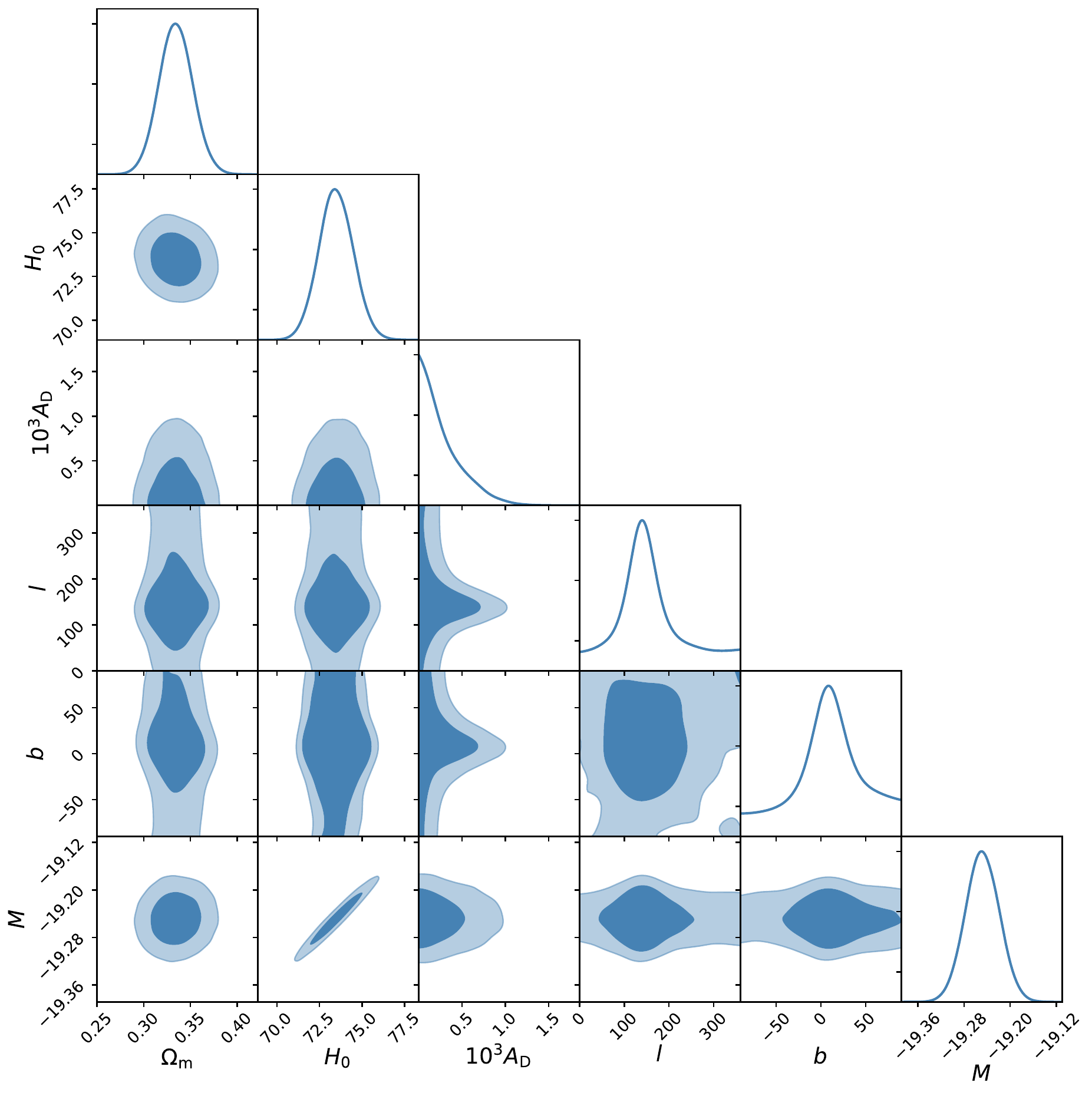}
		\caption{Full Pantheon+}
		\label{fig:contour_full}
	\end{subfigure}
	\begin{subfigure}[b]{0.49\textwidth}
		\centering
		\includegraphics[width=\textwidth,height=0.21\textheight,keepaspectratio]{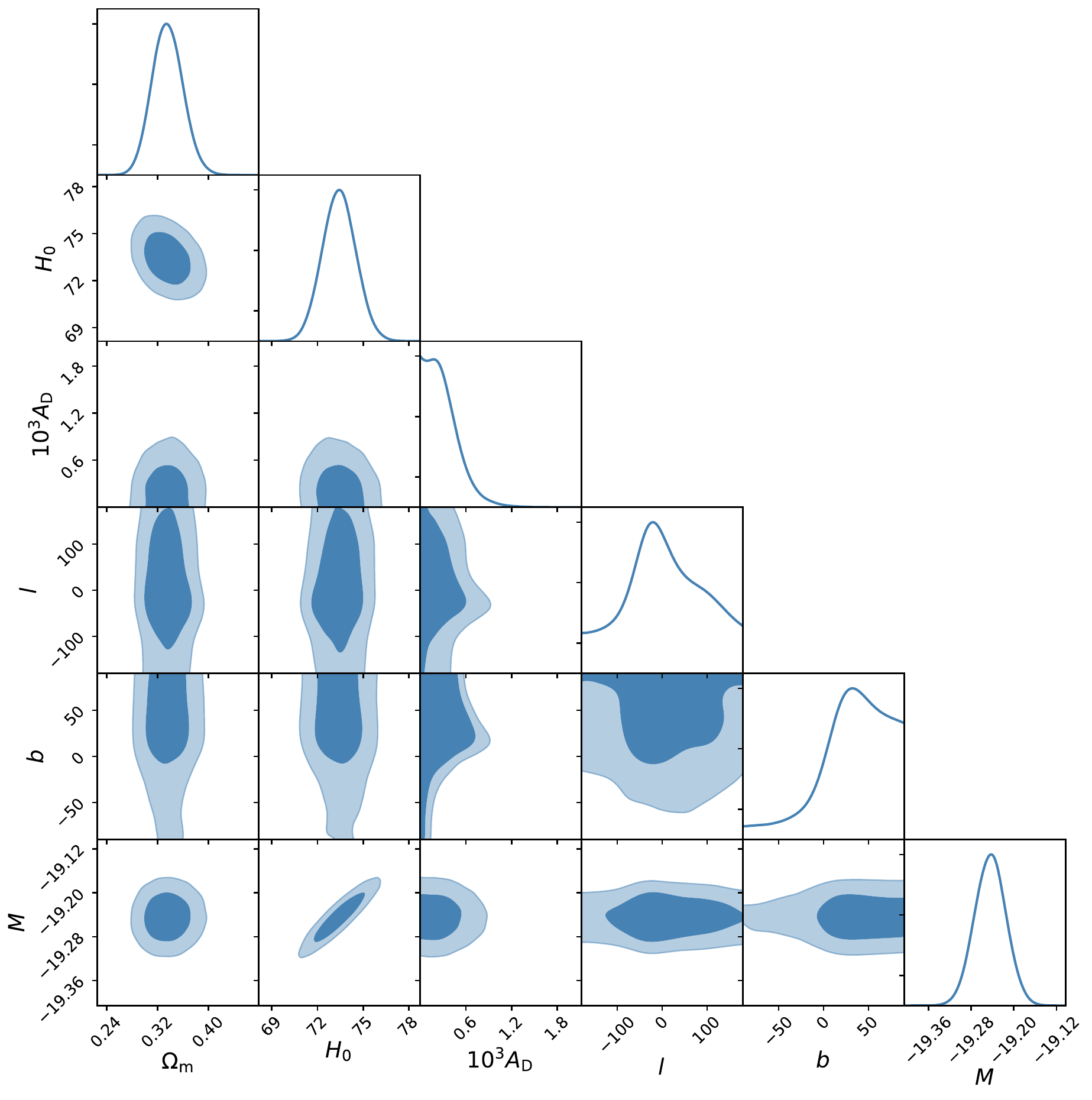}
		\caption{Excluding Low-$z$}
		\label{fig:contour_elowz}
	\end{subfigure}
	\hfill
	\begin{subfigure}[b]{0.49\textwidth}
		\centering
		\includegraphics[width=\textwidth,height=0.21\textheight,keepaspectratio]{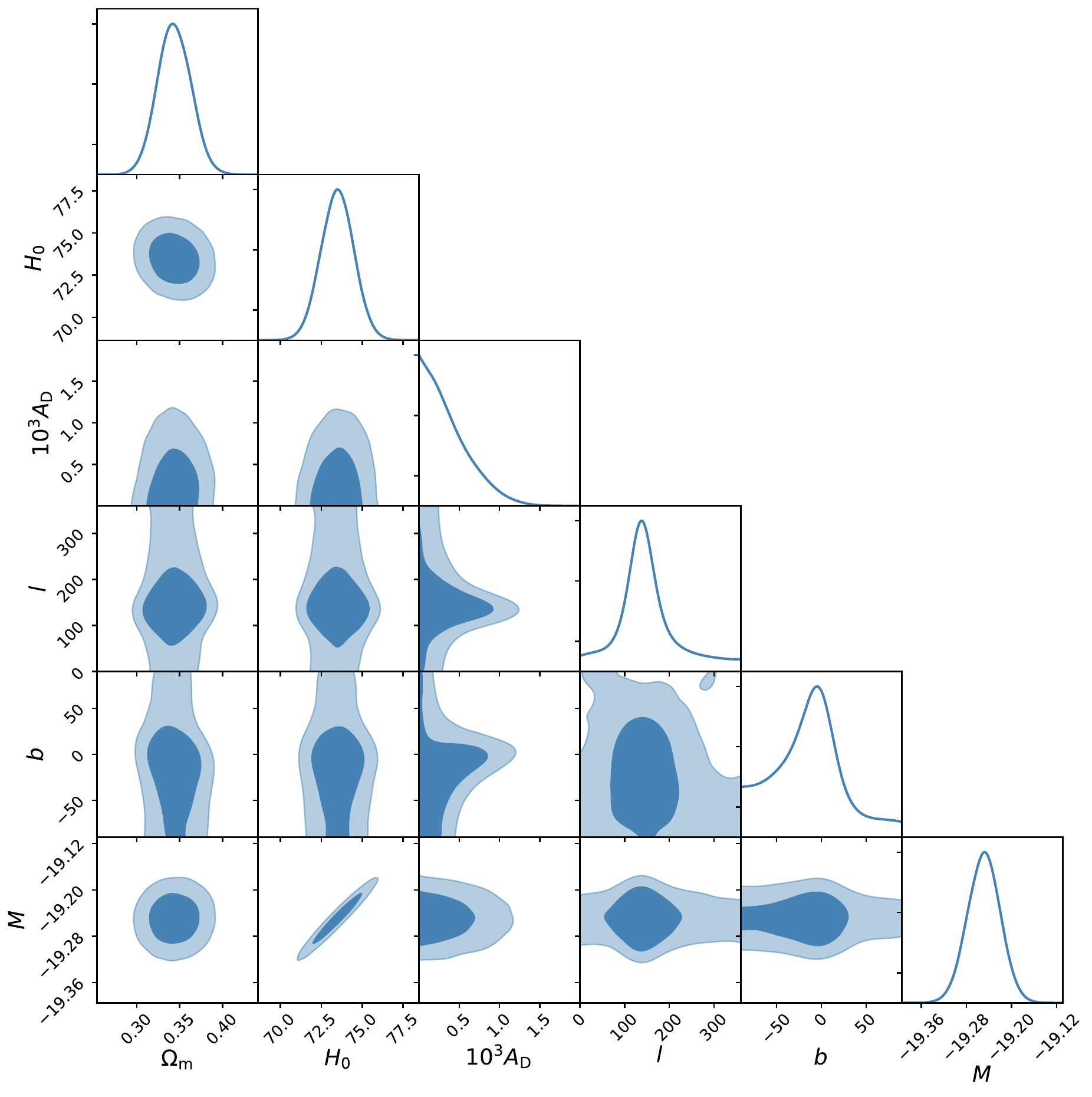}
		\caption{Excluding PS1}
		\label{fig:contour_ePS1}
	\end{subfigure}
    	\begin{subfigure}[b]{0.49\textwidth}
		\centering
		\includegraphics[width=\textwidth,height=0.21\textheight,keepaspectratio]{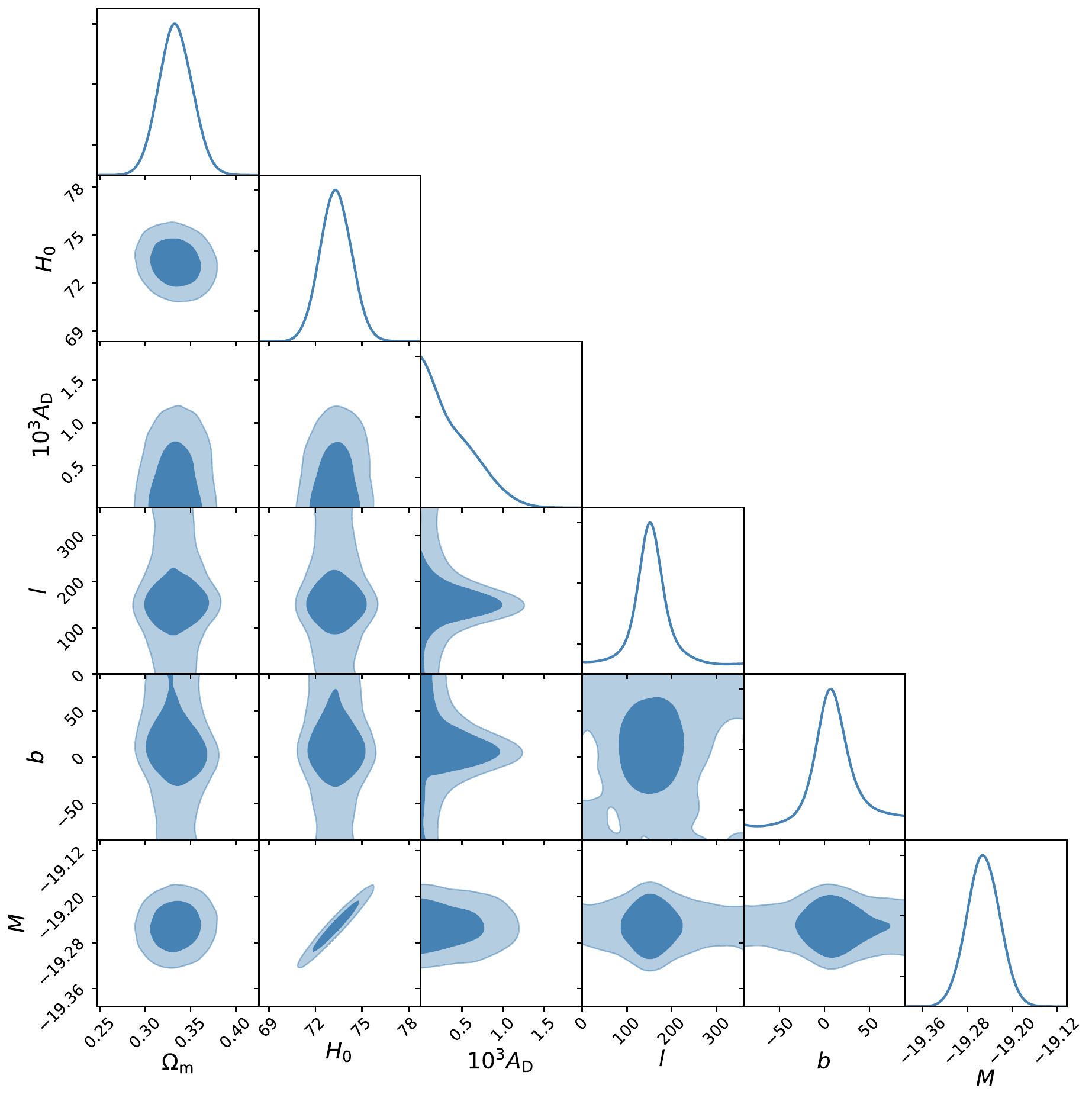}
		\caption{Excluding SDSS}
		\label{fig:contour_eSDSS}
	\end{subfigure}
	\begin{subfigure}[b]{0.49\textwidth}
		\centering
		\includegraphics[width=\textwidth,height=0.21\textheight,keepaspectratio]{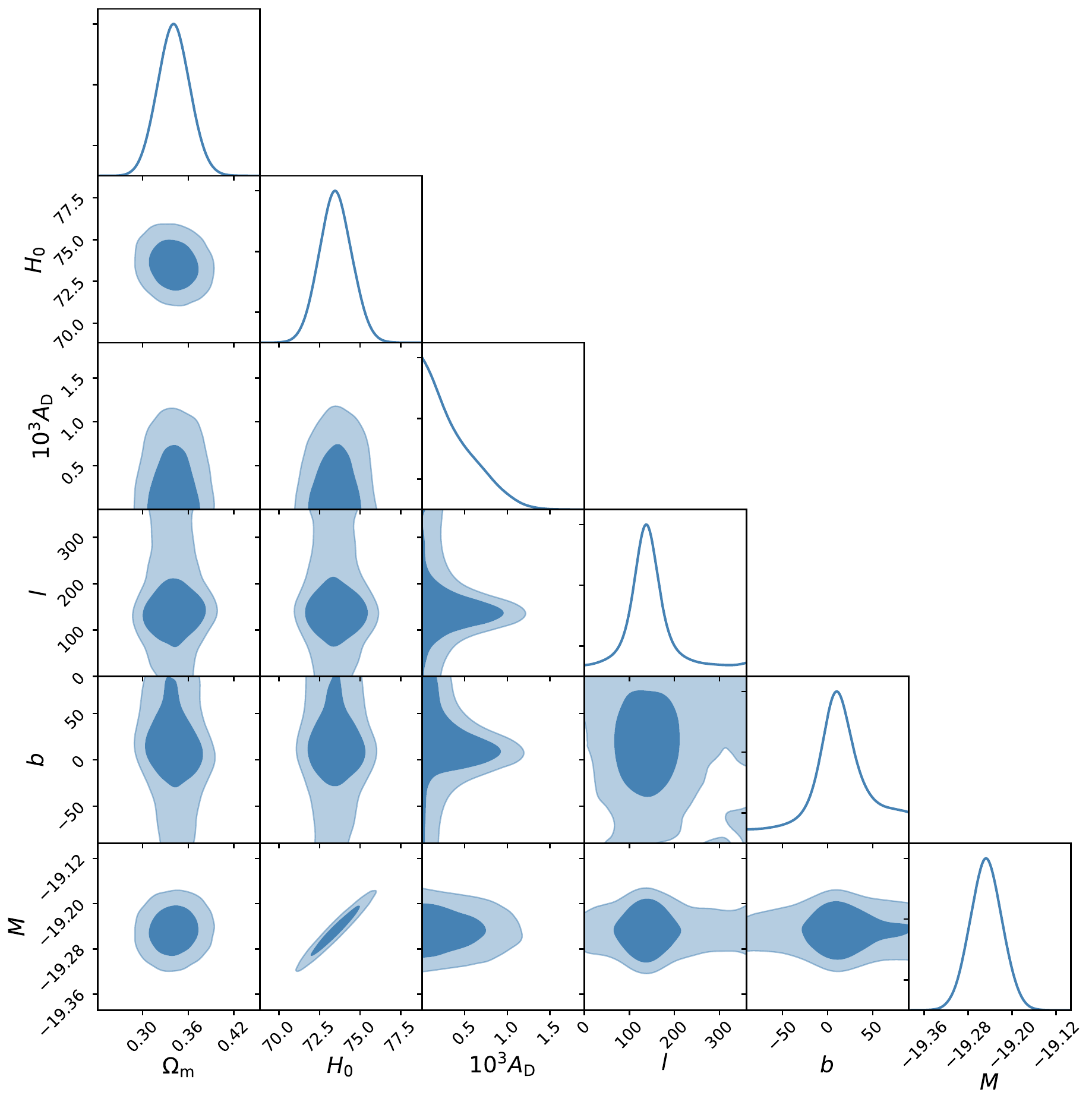}
		\caption{Excluding SNLS}
		\label{fig:contour_eSNLS}
	\end{subfigure}
    	\begin{subfigure}[b]{0.49\textwidth}
		\centering
		\includegraphics[width=\textwidth,height=0.21\textheight,keepaspectratio]{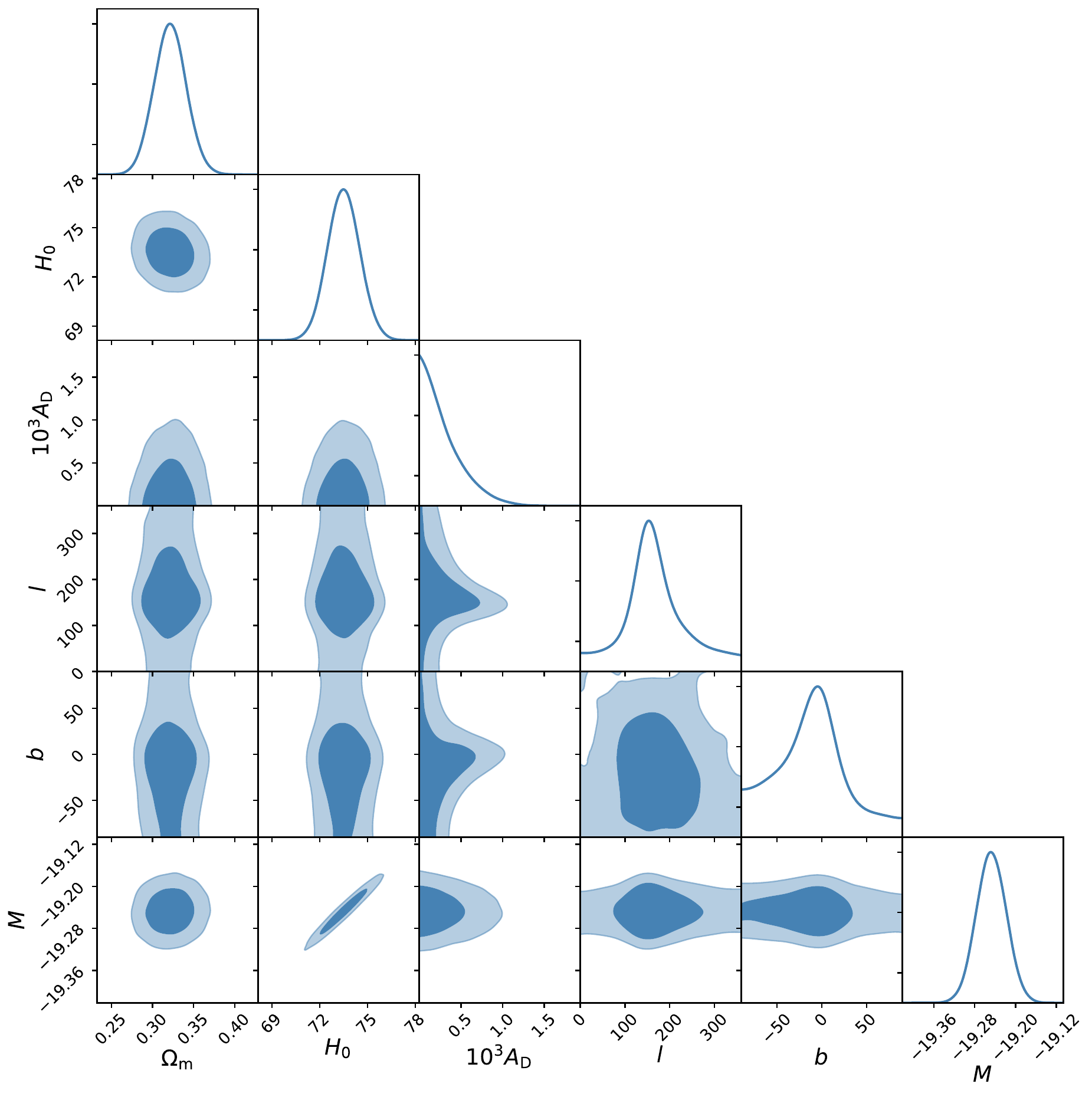}
		\caption{Excluding DES}
		\label{fig:contour_eDES}
	\end{subfigure}
	\hfill
	\begin{subfigure}[b]{0.49\textwidth}
		\centering
		\includegraphics[width=\textwidth,height=0.21\textheight,keepaspectratio]{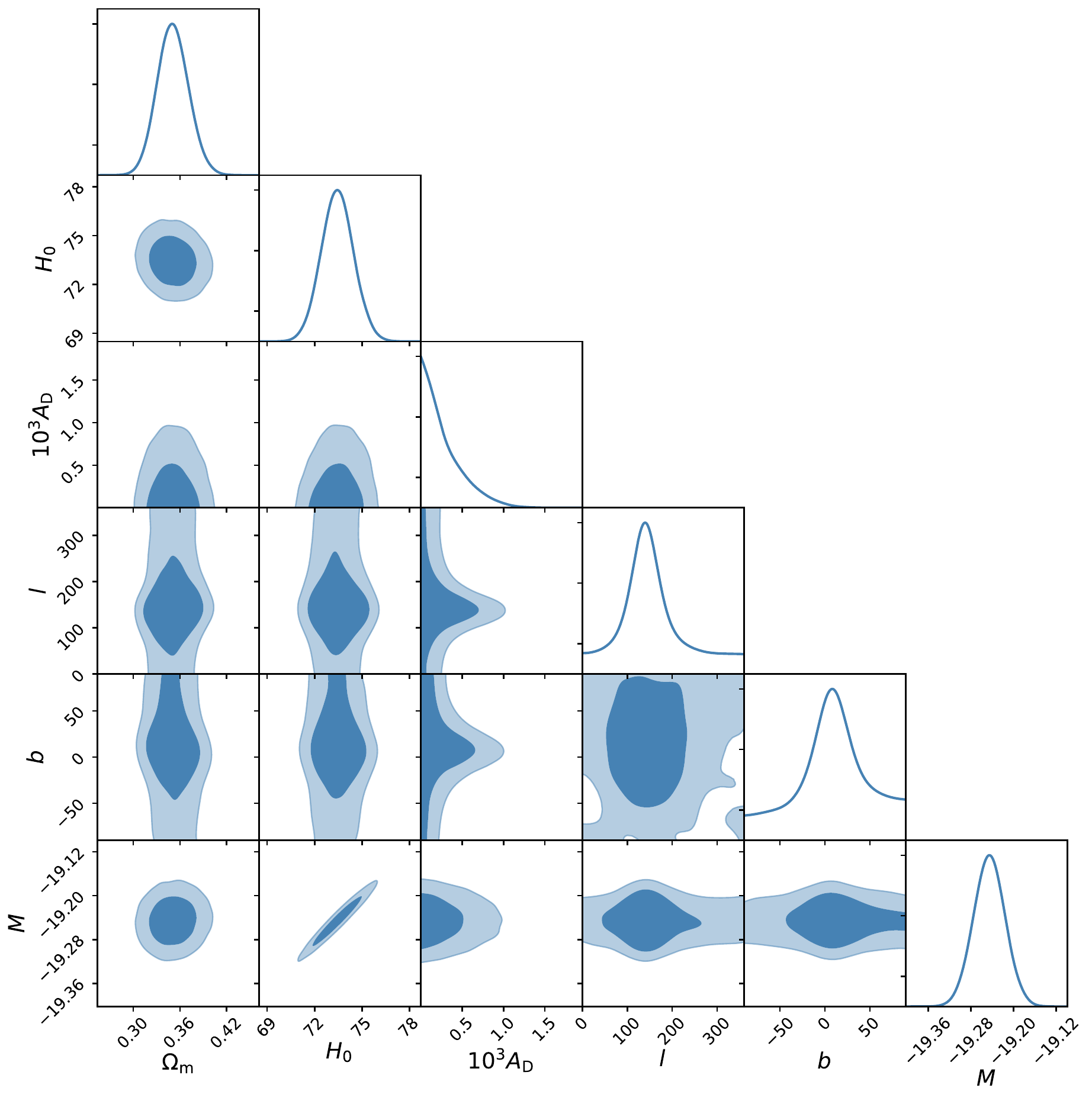}
		\caption{Excluding High-$z$}
		\label{fig:contour_eHighz}
	\end{subfigure}
	\caption{1D and 2D marginalized posterior distributions of the parameters for the full Pantheon+ sample, as well as Pantheon+ excluding the Low-$z$, PS1, SDSS, SNLS, DES, and High-$z$ subsamples, respectively.}
	\label{fig:contour_8}
\end{figure*}

To explore the full parameter space, we employ the Markov Chain Monte Carlo (MCMC) method implemented via the Python package \textit{emcee}\footnote{https://emcee.readthedocs.io/en/stable/} \citep{Foreman-Mackey:2012any}. We adopt flat priors for the free parameters as follows: $\Omega_\mathrm{m}\in[0, 1]$, $H_0\in[50, 90]$, $A_\mathrm{D}\in[0, 0.5]$, $l\in[0^\circ, 360^\circ]$, $b\in[-90^\circ, 90^\circ]$, and $M\in[-20.5, -18]$. However, for the `Pantheon+ excluding Low-$z$' combination, the flat prior for $l$ is set to $[-180^\circ, 180^\circ]$. Figure \ref{fig:contour_8} displays the 1D and 2D marginalized posterior distributions of the free parameters, with the best-fit values and their 68\% highest-density intervals compiled in Table \ref{table:DF}.

\begin{table*}[!htbp]
	\centering
	\caption{Best-fit values of the parameters with 68\% highest-density intervals for: (i) the full Pantheon+ sample, (ii) Pantheon+ excluding the Low-$z$, PS1, SDSS, SNLS, DES, and High-$z$ subsamples, respectively, (iii) the Low-$z$ subsample only, and (iv) the combined subset of surveys 5, 56, 63, and 150. For parameters with only one-sided limits, the 95\% confidence level is reported instead. Here, $N$ represents the total number of SNe Ia. The numbers in parentheses denote the additionally retained Cepheid-hosting SNe Ia. For the `Pantheon+ excluding Low-$z$' case, the Galactic longitude $l$ is converted to the range $[0^\circ, 360^\circ]$.}
	\label{table:DF}
	\setlength{\tabcolsep}{2.85mm}
	\renewcommand{\arraystretch}{1.3}
	\resizebox{\textwidth}{!}{%
		\begin{tabular}{lccccccc}
			\hline
			Sample & $N$ & $\Omega_{\mathrm{m}}$ & $H_0$\,[$\mathrm{km\,s^{-1}\,Mpc^{-1}}$]  & $10^3A_{\mathrm{D}}$ & $l\,[^\circ]$ & $b\,[^\circ]$ & $M$ \\
			\hline
			Full Pantheon+
            & $1701$
			& $0.333_{-0.017}^{+0.019}$
            & $73.458_{-1.040}^{+0.975}$
            & $<0.764$
            & $138.39_{-57.98}^{+70.43}$
            & $8.53_{-33.72}^{+42.67}$
            & $-19.252_{-0.025}^{+0.033}$ \\

			Excluding Low-$z$
            & $983\,(77)$
            & $0.334_{-0.025}^{+0.024}$
            & $73.571_{-1.271}^{+0.937}$
            & $<0.703$
            & $334.02_{-45.37}^{+138.16}$
            & $30.63_{-19.36}^{+50.78}$
            & $-19.241_{-0.033}^{+0.026}$ \\

			Excluding PS1
            & $1432$
            & $0.343_{-0.019}^{+0.020}$
            & $73.492_{-1.035}^{+0.983}$
            & $<0.927$
            & $136.72_{-50.84}^{+57.52}$
            & $-4.78_{-48.10}^{+25.60}$
            & $-19.250_{-0.028}^{+0.030}$ \\

			Excluding SDSS
            & $1380$
            & $0.333_{-0.019}^{+0.019}$
            & $73.231_{-0.934}^{+1.093}$
            & $<0.968$
            & $150.57_{-43.93}^{+47.18}$
            & $6.46_{-27.05}^{+33.64}$
            & $-19.255_{-0.025}^{+0.033}$ \\

			Excluding SNLS
            & $1541$
            & $0.341_{-0.021}^{+0.021}$
            & $73.400_{-0.921}^{+1.093}$
            & $<0.941$
            & $136.80_{-44.45}^{+49.90}$
            & $9.64_{-26.51}^{+38.14}$
            & $-19.248_{-0.029}^{+0.029}$ \\

			Excluding DES
            & $1498$
            & $0.322_{-0.020}^{+0.019}$
            & $73.499_{-0.971}^{+1.048}$
            & $<0.779$
            & $151.04_{-47.60}^{+79.48}$
            & $-2.19_{-51.53}^{+25.60}$
            & $-19.253_{-0.024}^{+0.035}$ \\

			Excluding High-$z$
            & $1671$
            & $0.351_{-0.021}^{+0.019}$
            & $73.484_{-1.054}^{+0.985}$
            & $<0.766$
            & $139.10_{-62.24}^{+65.23}$
            & $7.20_{-35.02}^{+43.94}$
            & $-19.248_{-0.025}^{+0.034}$ \\

			Low-$z$
            & $718$
            & $0.438_{-0.252}^{+0.245}$
            & $73.130_{-1.124}^{+1.103}$
            & $0.952_{-0.403}^{+0.454}$
            & $149.77_{-20.96}^{+24.21}$
            & $-12.20_{-20.93}^{+18.59}$
            & $-19.247_{-0.031}^{+0.028}$ \\

			5+56+63+150
            & $359\,(46)$
            & $0.287_{-0.220}^{+0.272}$
            & $72.994_{-1.122}^{+1.192}$
            & $1.730_{-0.715}^{+0.554}$
            & $153.05_{-16.56}^{+15.53}$
            & $-1.25_{-16.04}^{+11.93}$
            & $-19.251_{-0.034}^{+0.025}$ \\
			\hline
		\end{tabular}%
	}
\end{table*}

\begin{figure}
	\begin{center}
		\includegraphics[width=8cm]{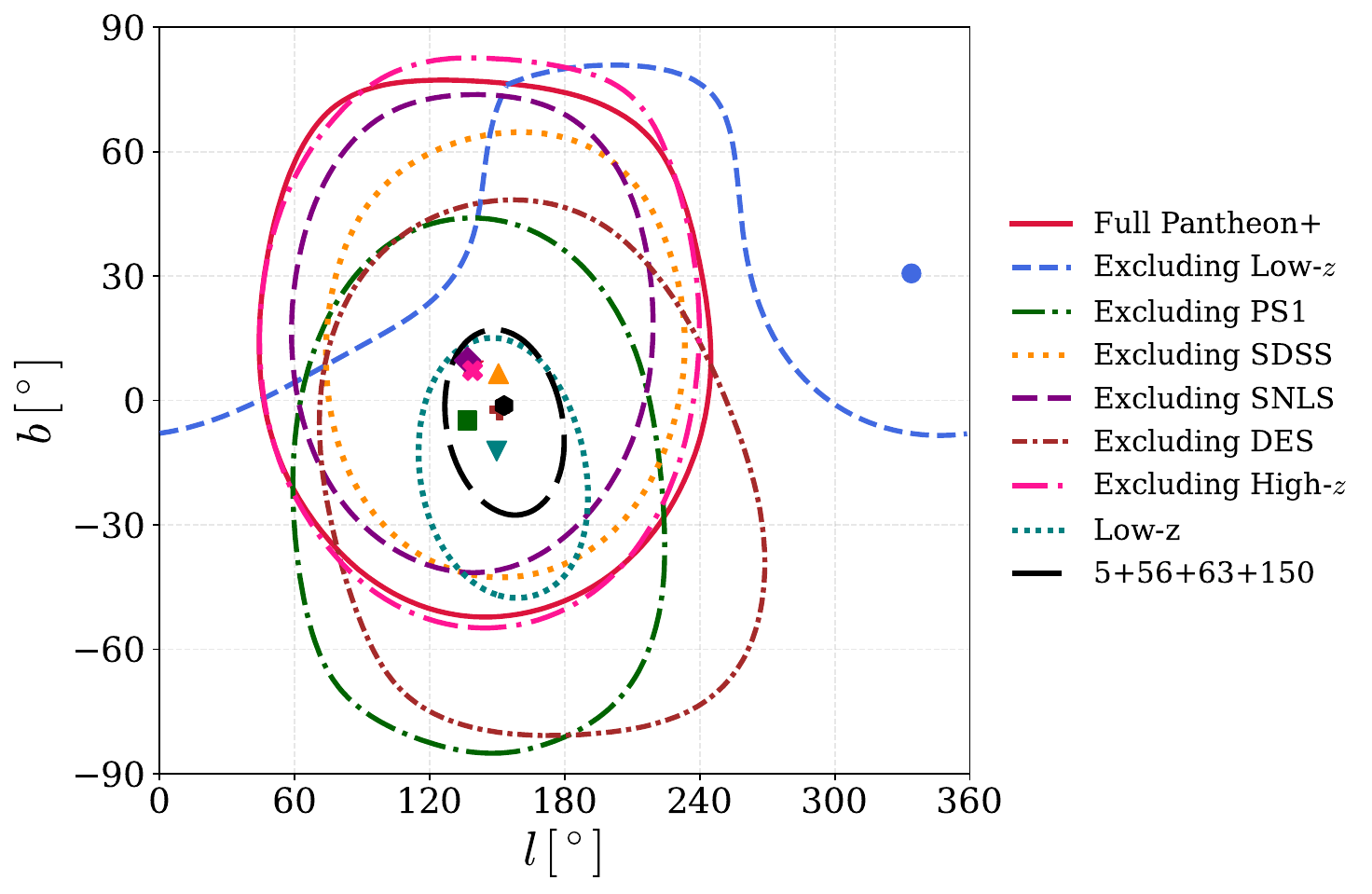}
		\caption{2D marginalized posterior distributions of the Galactic coordinates $l$ and $b$ for: (i) the full Pantheon+ sample, (ii) Pantheon+ excluding the Low-$z$, PS1, SDSS, SNLS, DES, and High-$z$ subsamples, respectively, (iii) the Low-$z$ subsample only, and (iv) the combined subset of surveys 5, 56, 63, and 150.}
		\label{fig:lb_contour}
	\end{center}
\end{figure}

For the full Pantheon+ sample, we obtain $\Omega_{\mathrm{m}}=0.333^{+0.019}_{-0.017}$, $H_0=73.458^{+0.975}_{-1.040}$, and $M=-19.252^{+0.033}_{-0.025}$. The dipole amplitude is constrained only by a 95\% confidence level (CL) upper limit of $A_{\mathrm{D}}<0.764\times10^{-3}$, indicating that the dipole anisotropy in the full Pantheon+ sample is weak. The associated direction is $(l,b)=(138.39^\circ,8.53^\circ)$. To evaluate the influence of individual subsamples on the overall anisotropy, we systematically exclude each Pantheon+ subsample in turn. Excluding the PS1, SDSS, SNLS, DES, or High-$z$ subsamples individually yields results that are broadly consistent with those of the full Pantheon+ sample. In these cases, the dipole amplitude remains constrained only by an upper limit, and the dipole directions remain closely aligned with the original preferred direction. Conversely, when the Low-$z$ subsample is excluded, the dipole direction shifts significantly to $(l,b)=(334.02^\circ, 30.63^\circ)$, representing an angular separation of $138.07^\circ$ from the full-sample result. Figure \ref{fig:lb_contour} displays the 2D marginalized posterior distributions of the Galactic coordinates $l$ and $b$. Specifically, the dipole directions obtained from the full sample and the cases excluding the PS1, SDSS, SNLS, DES, or High-$z$ subsamples remain consistent with one another. Conversely, when the Low-$z$ subsample is excluded, the dipole direction shifts beyond the $1\sigma$ uncertainties of the aforementioned cases.

\begin{figure*}[!htbp]
	\centering
	\begin{subfigure}[b]{0.49\textwidth}
		\centering
		\includegraphics[width=\textwidth]{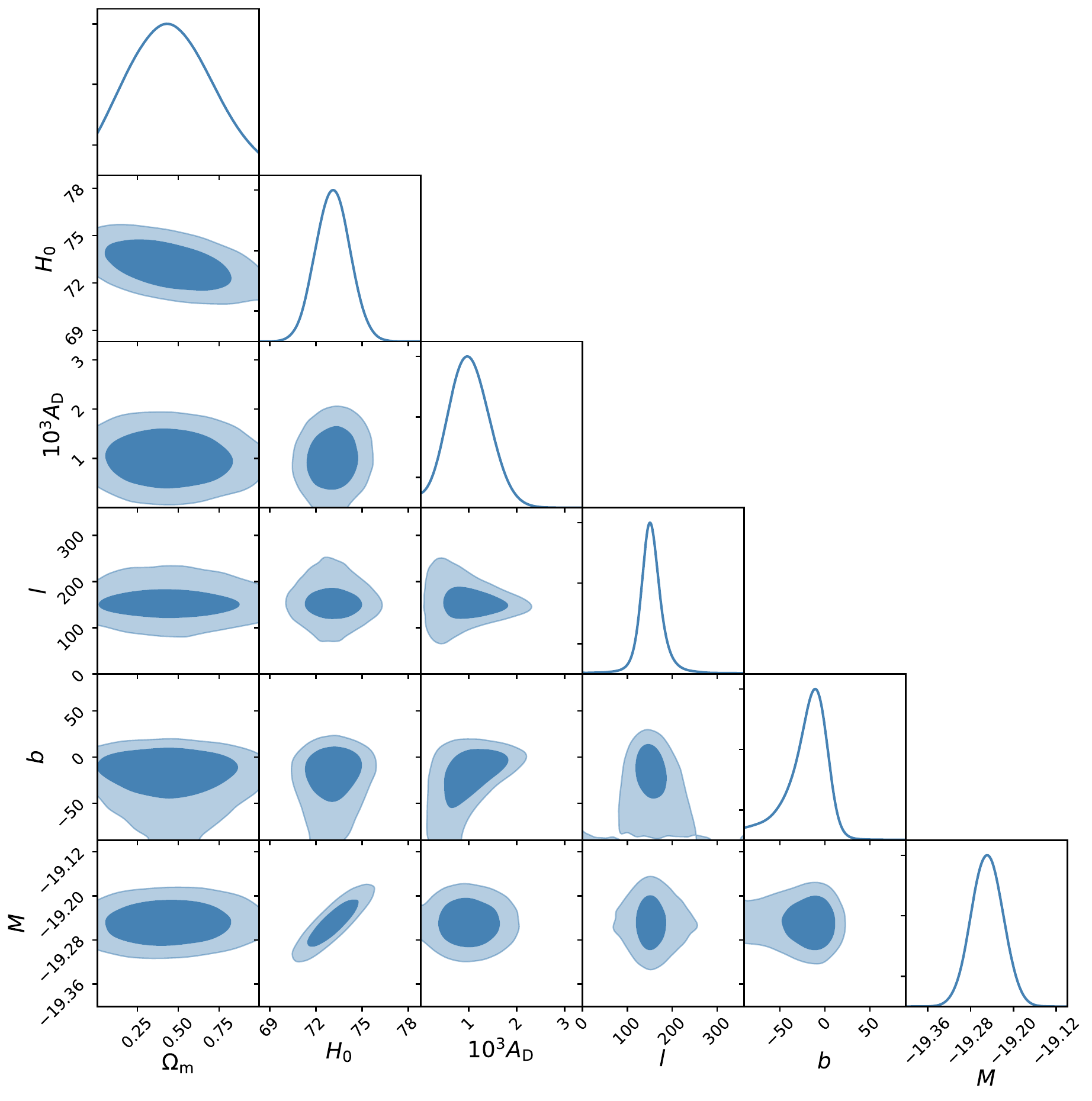}
		\caption{Low-$z$}
		\label{fig:contour_lowz}
	\end{subfigure}
    	\begin{subfigure}[b]{0.49\textwidth}
		\centering
		\includegraphics[width=\textwidth]{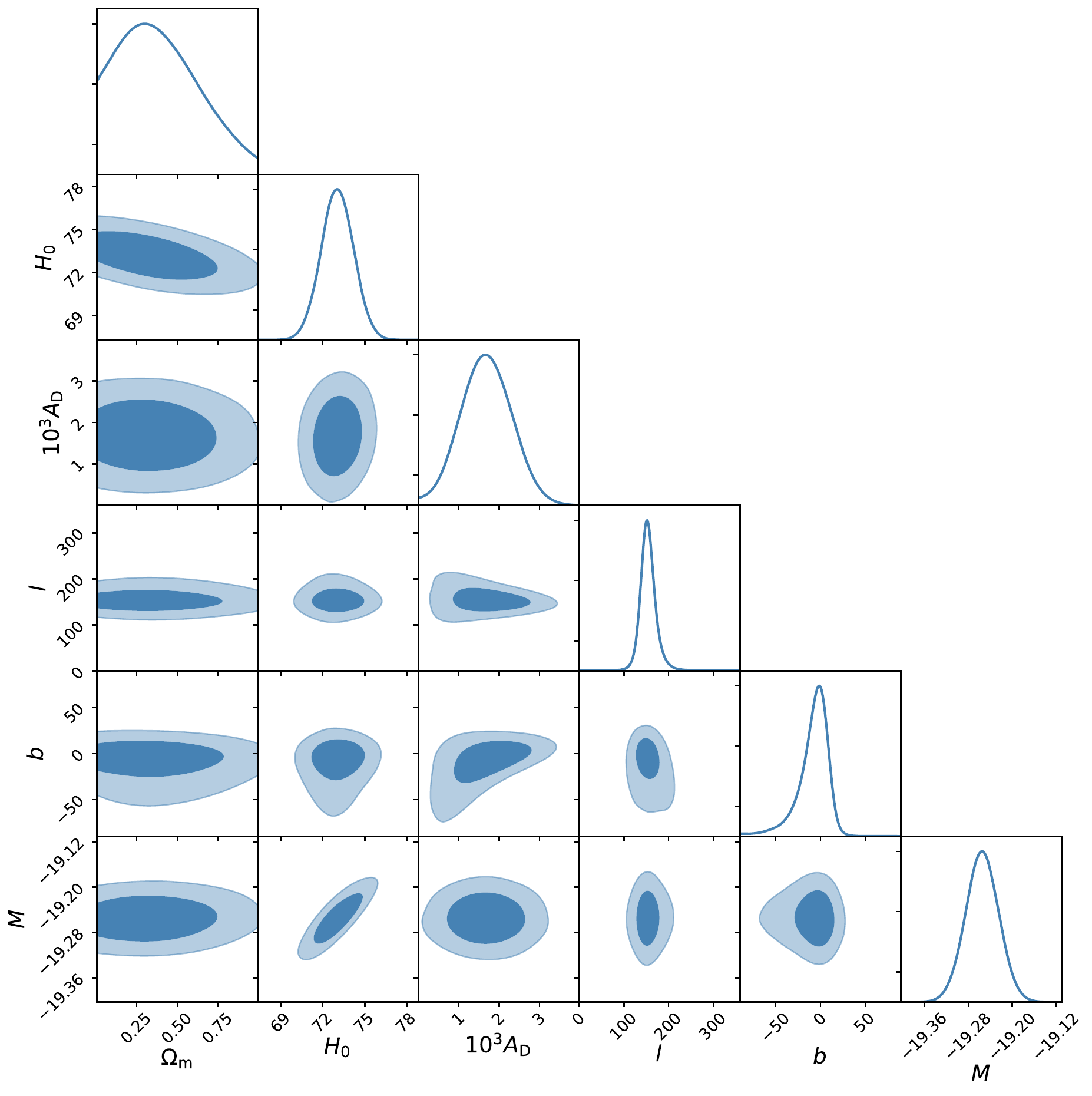}
		\caption{5+56+63+150}
		\label{fig:contour_survey}
	\end{subfigure}
	\caption{1D and 2D marginalized posterior distributions of the parameters for the Low-$z$ subsample and the combined subset of surveys 5, 56, 63, and 150, respectively.}
	\label{fig:contour_2}
\end{figure*}

Given the significant impact of the Low-$z$ subsample, we perform a separate analysis on this subset. The dipole amplitude is found to be $A_{\mathrm{D}}=0.952^{+0.454}_{-0.403}\times10^{-3}$ with a dipole direction of $(l,b)=(149.77^\circ,-12.20^\circ)$. Unlike the full sample, the Low-$z$ subsample yields a marginally non-zero dipole amplitude. This direction is closely aligned with that of the full Pantheon+ sample, with an angular separation of only $23.62^\circ$. Furthermore, we investigate the individual surveys constituting the Low-$z$ subsample. The dipole signal is found to be predominantly associated with surveys 5, 56, 63, and 150. For this combined subset, we obtain $A_{\mathrm{D}}=1.730_{-0.715}^{+0.554}\times10^{-3}$ with a dipole direction of $(l,b)=(153.05^\circ,-1.25^\circ)$. This direction is closely aligned with those of the Low-$z$ subsample and the full Pantheon+ sample, with an angular separation of only $11.42^\circ$ and $15.45^\circ$, respectively. For the Low-$z$ subsample and the combined subset of surveys 5, 56, 63, and 150, we show the 1D and 2D marginalized posterior distributions of the free parameters (Figure \ref{fig:contour_2}) and compile the corresponding best-fit values along with their 68\% highest-density intervals (Table \ref{table:DF}). Figure \ref{fig:lb_contour} displays the 2D marginalized posterior distributions of the Galactic coordinates $l$ and $b$. Notably, for both datasets, the uncertainties on the Galactic coordinates $l$ and $b$ are significantly reduced.

\begin{figure}[!htbp]
	\centering
	\begin{subfigure}[b]{0.49\textwidth}
		\centering
		\includegraphics[width=\textwidth]{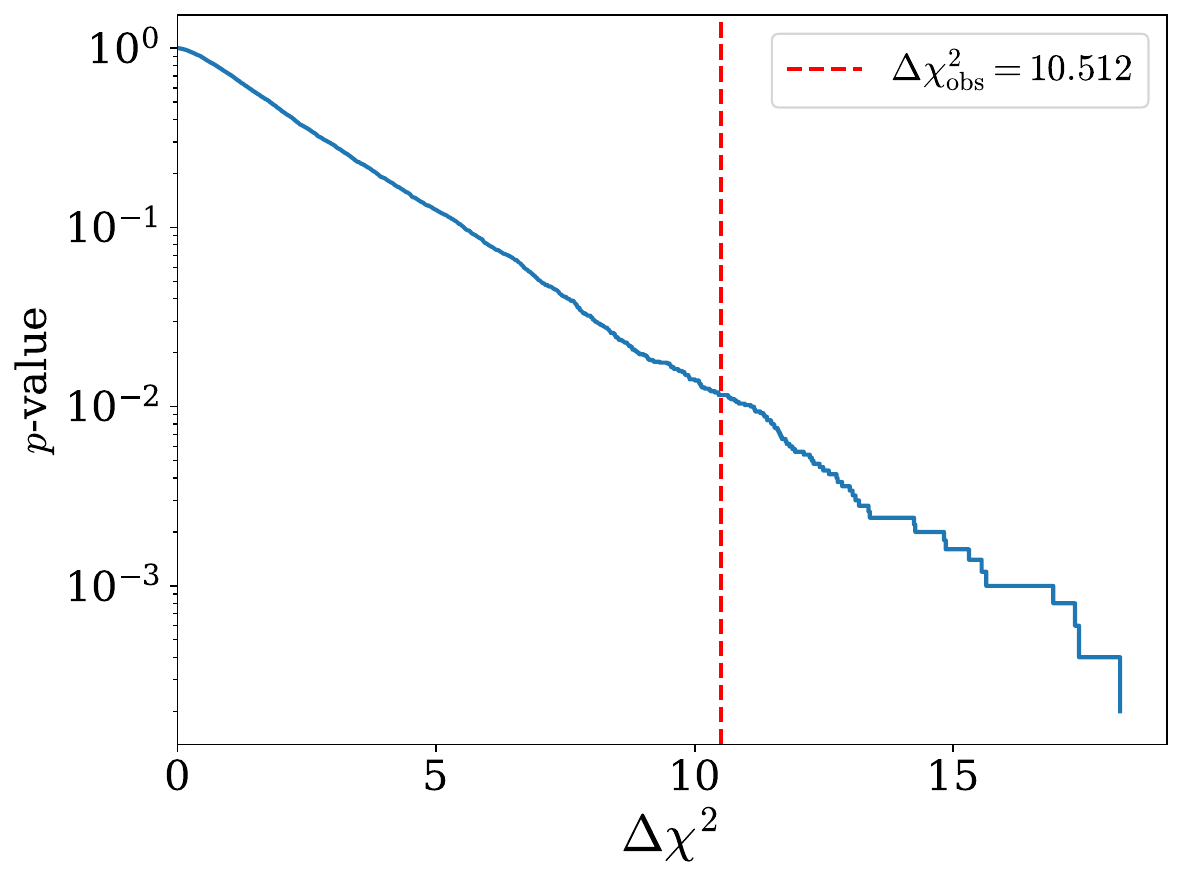}
		\caption{Low-$z$}
		\label{fig:survival_lowz}
	\end{subfigure}
    \begin{subfigure}[b]{0.49\textwidth}
		\centering
		\includegraphics[width=\textwidth]{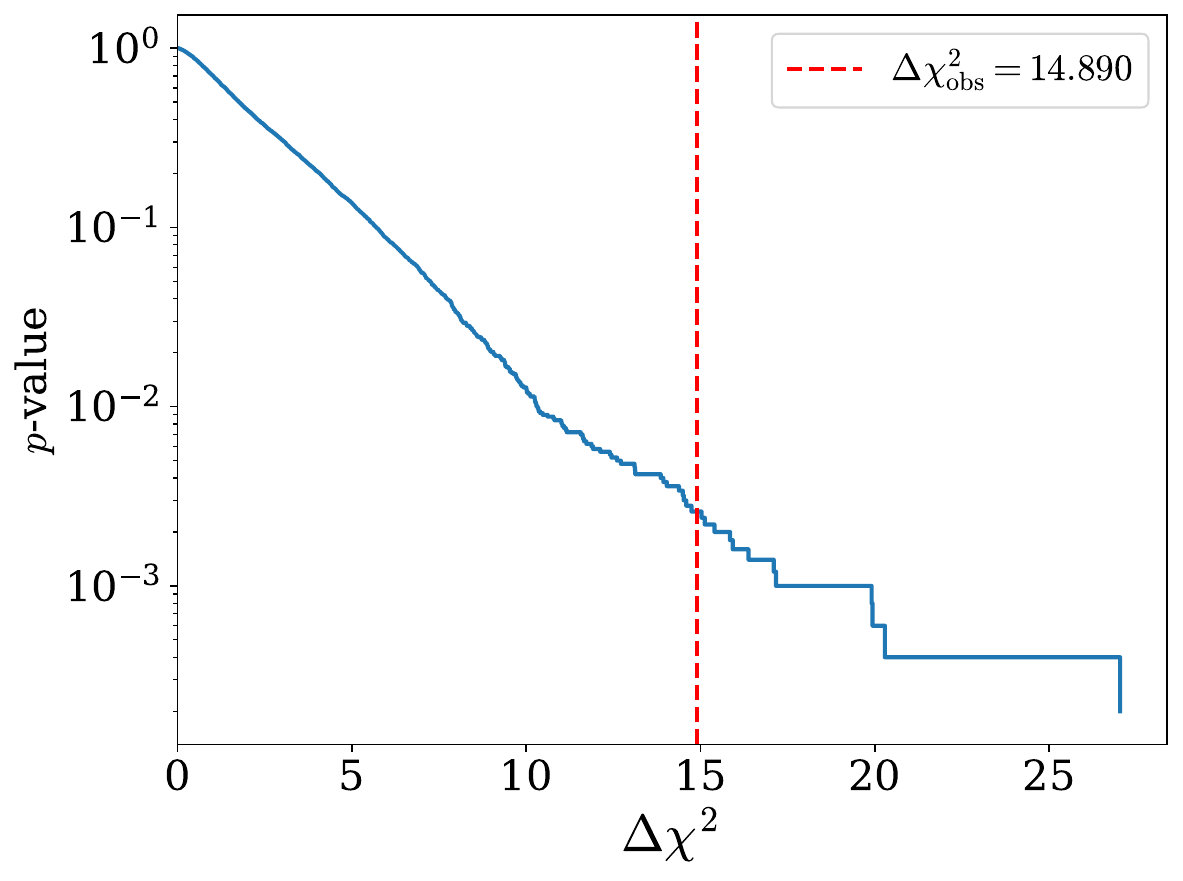}
		\caption{5+56+63+150}
		\label{fig:survival_survey}
	\end{subfigure}
	\caption{Empirical probability of obtaining $\Delta\chi^2_{\mathrm{mock}} \geq \Delta\chi^2$ from the isotropic mock samples. The horizontal and vertical axes represent the threshold $\Delta\chi^2$ and the fraction of mock samples satisfying this condition, respectively. The vertical dashed line indicates the observed value $\Delta\chi^2_{\mathrm{obs}}$ from the actual data. The upper and lower panels correspond to the Low-$z$ subsample and the combined subset of surveys 5, 56, 63, and 150, respectively.}
	\label{fig:survival}
\end{figure}

We conduct an isotropic mock analysis to evaluate the statistical significance of the dipole signals in both the Low-$z$ subsample and the combined subset of surveys 5, 56, 63, and 150. Initially, the selected sample is fitted with the flat $\Lambda\mathrm{CDM}$ model, which corresponds to fixing the dipole amplitude to $A_{\mathrm{D}}=0$. Using the best-fit isotropic parameters $(\Omega_{\mathrm{m}}, H_0, M)$, we generate $N_{\mathrm{mock}}=5000$ mock samples. Gaussian fluctuations from the full covariance matrix are then added to the isotropic predictions. The redshifts, sky positions, calibrator flags, and covariance matrix remain identical to those of the actual data. For each mock sample, we minimize the $\chi^2$ of both the isotropic and dipole models to obtain the corresponding $\Delta\chi^2_{\mathrm{mock}}$. The observed value $\Delta\chi^2_{\mathrm{obs}}$ is computed in an identical manner from the actual dataset. The empirical $p$-value is then defined as:
\begin{equation}
    p_{\Delta\chi^2} =
    \frac{
    N\left(\Delta\chi^2_{\mathrm{mock}}
    \ge
    \Delta\chi^2_{\mathrm{obs}}\right)+1
    }{
    N_{\mathrm{mock}}+1
    }.
\end{equation}
The corresponding Gaussian-equivalent one-sided significance is obtained from this $p$-value. For the Low-$z$ subsample, we find $p_{\Delta\chi^2} = 0.0116$, corresponding to a Gaussian-equivalent one-sided significance of $2.27\sigma$. Similarly, the combined subset of surveys 5, 56, 63, and 150 yields $p_{\Delta\chi^2} = 0.0026$, which translates to a significance of $2.79\sigma$. Figure \ref{fig:survival} displays the $p$-value derived from the isotropic mock realizations.

\section{Hemisphere comparison method and result}\label{HC}
The HC method is executed through the following procedure:

(1) First, a random direction vector $\hat{\boldsymbol{n}}(l, b)$ is generated in Galactic coordinates, where $l$ and $b$ denote the Galactic longitude and latitude, respectively. This vector $\hat{\boldsymbol{n}}$ serves to bisect the celestial sphere into two opposing ``up'' and ``down'' hemispheres.

(2) Next, based on the spatial coordinates of SNe Ia, the Pantheon+ sample is separated into two subsets corresponding to these respective hemispheres.

(3) For each hemispherical subset, we determine the best-fit value of $\Omega_\mathrm{m}$ and its associated $1\sigma$ uncertainty under the assumption of a flat $\Lambda\mathrm{CDM}$ model. The anisotropy level (AL) is defined as:
\begin{equation}
	\mathrm{AL}=2 \times \frac{\Omega_{\mathrm{m}, u}-\Omega_{\mathrm{m}, d}}{\Omega_{\mathrm{m}, u}+\Omega_{\mathrm{m}, d}},
\end{equation}
where the subscripts $u$ and $d$ refer to the ``up'' and ``down'' hemispheres, respectively. The corresponding $1\sigma$ uncertainty of the AL is calculated as:
\begin{equation}
	\sigma_{\mathrm{AL}}=\frac{\sqrt{\sigma_{\Omega_{\mathrm{m}, u}}^{2}+\sigma_{\Omega_{\mathrm{m}, d}}^{2}}}{\Omega_{\mathrm{m}, u}+\Omega_{\mathrm{m}, d}}.
\end{equation}

(4) This process is repeated over a sufficiently large number of random orientations to identify the maximum anisotropy level ($\mathrm{AL}_{\mathrm{max}}$) and its associated direction.

\begin{figure*}[!htbp]
	\centering
	\begin{subfigure}[b]{0.49\textwidth}
		\centering
		\includegraphics[width=\textwidth,height=0.21\textheight,keepaspectratio]{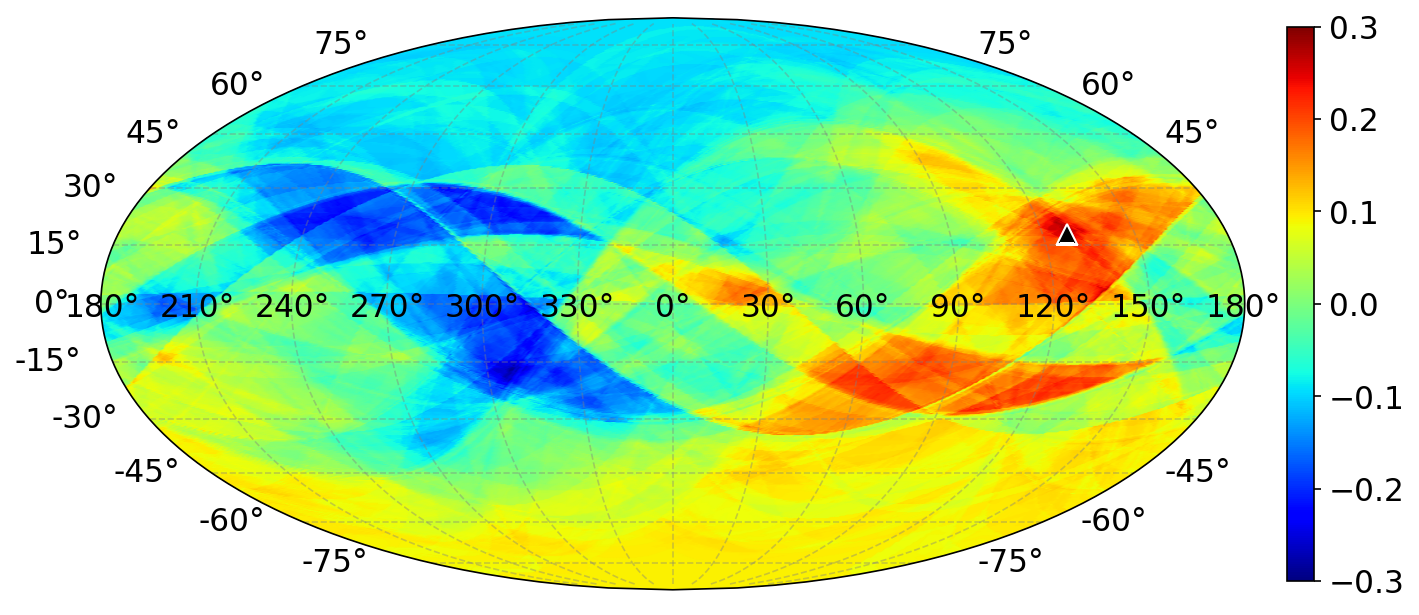}
		\caption{Full Pantheon+}
		\label{fig:HC_full}
	\end{subfigure}
	\begin{subfigure}[b]{0.49\textwidth}
		\centering
		\includegraphics[width=\textwidth,height=0.21\textheight,keepaspectratio]{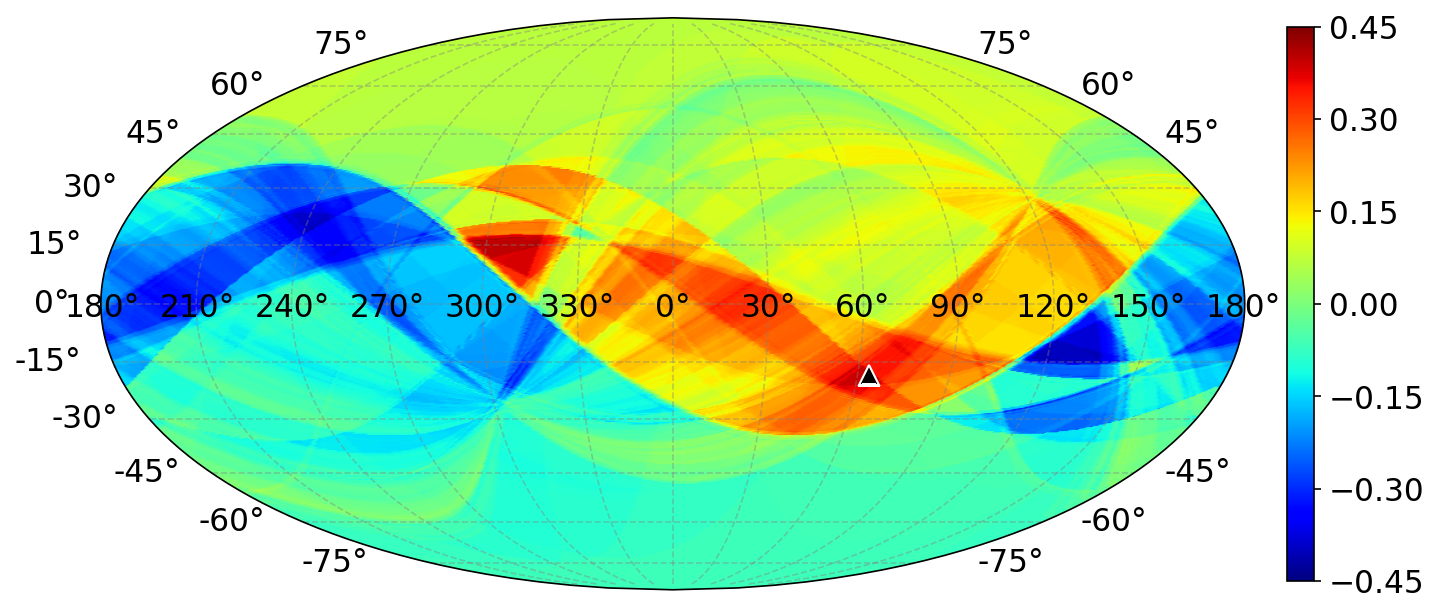}
		\caption{Excluding Low-$z$}
		\label{fig:HC_elowz}
	\end{subfigure}
	\hfill
	\begin{subfigure}[b]{0.49\textwidth}
		\centering
		\includegraphics[width=\textwidth,height=0.21\textheight,keepaspectratio]{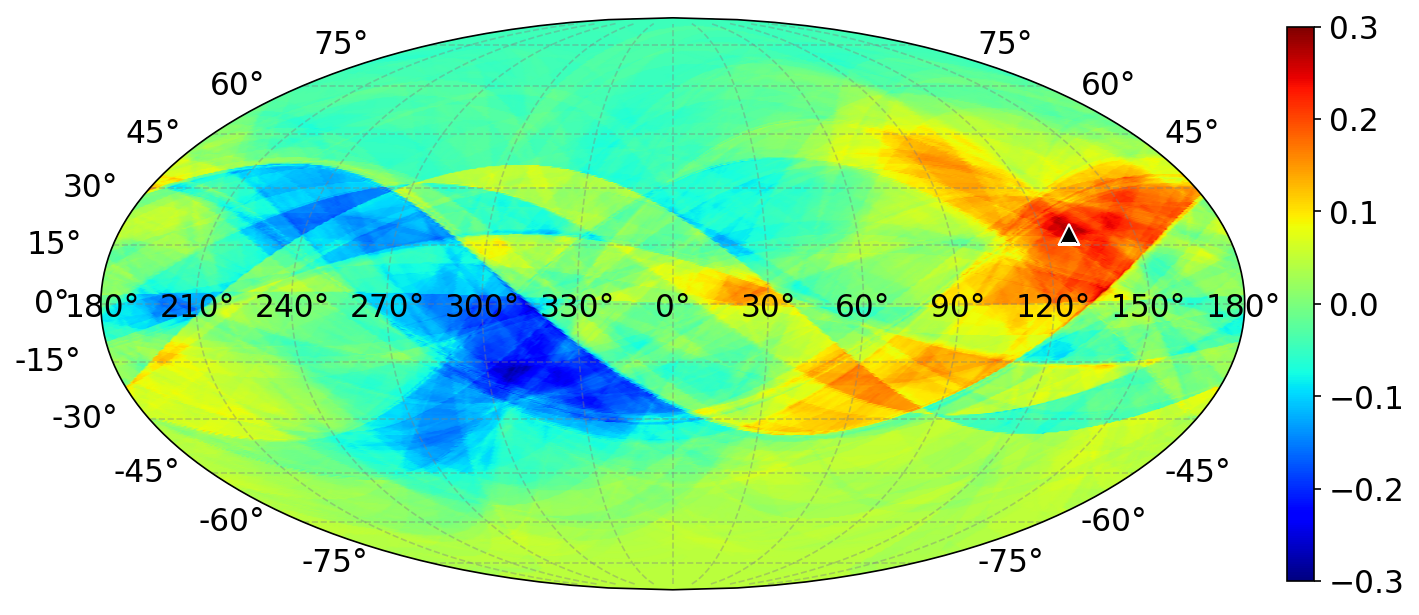}
		\caption{Excluding PS1}
		\label{fig:HC_ePS1}
	\end{subfigure}
    	\begin{subfigure}[b]{0.49\textwidth}
		\centering
		\includegraphics[width=\textwidth,height=0.21\textheight,keepaspectratio]{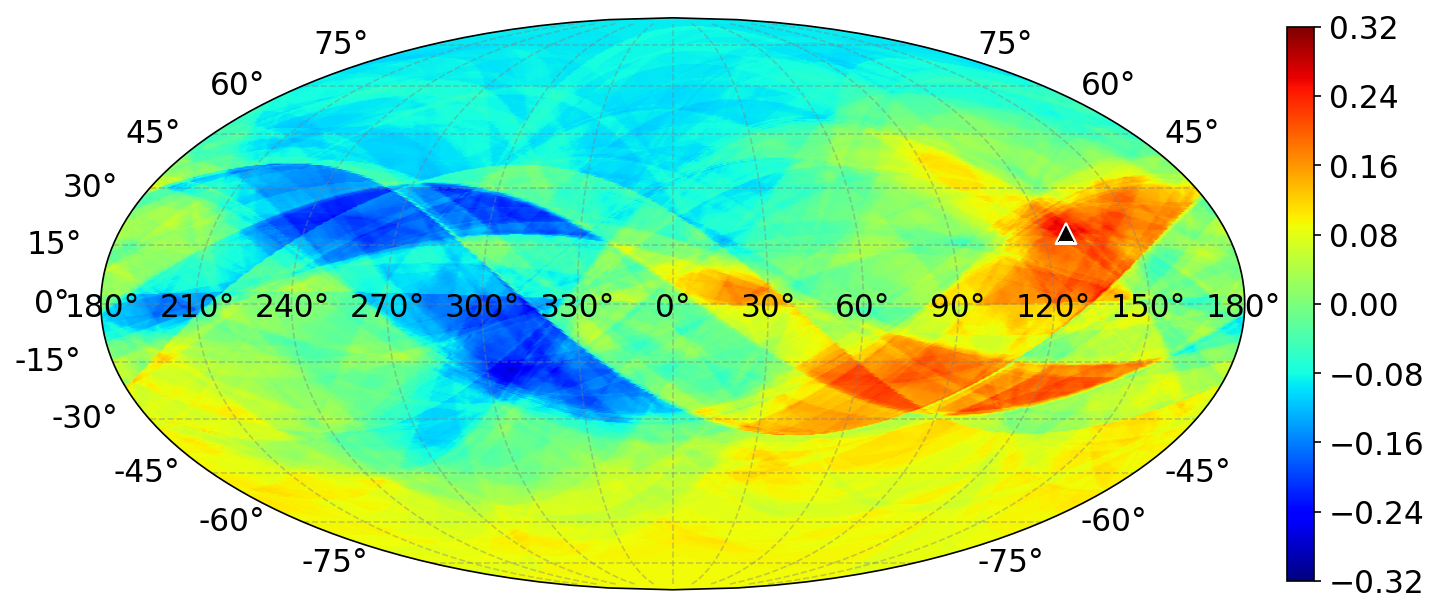}
		\caption{Excluding SDSS}
		\label{fig:HC_eSDSS}
	\end{subfigure}
	\begin{subfigure}[b]{0.49\textwidth}
		\centering
		\includegraphics[width=\textwidth,height=0.21\textheight,keepaspectratio]{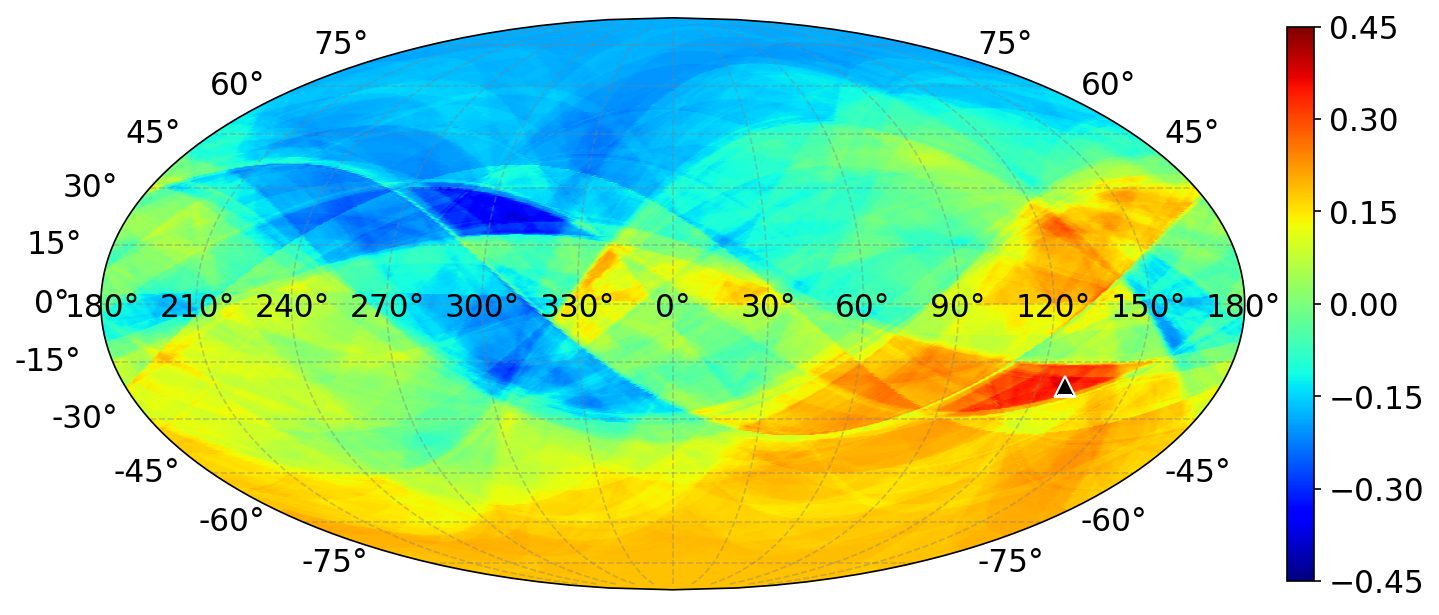}
		\caption{Excluding SNLS}
		\label{fig:HC_eSNLS}
	\end{subfigure}
    	\begin{subfigure}[b]{0.49\textwidth}
		\centering
		\includegraphics[width=\textwidth,height=0.21\textheight,keepaspectratio]{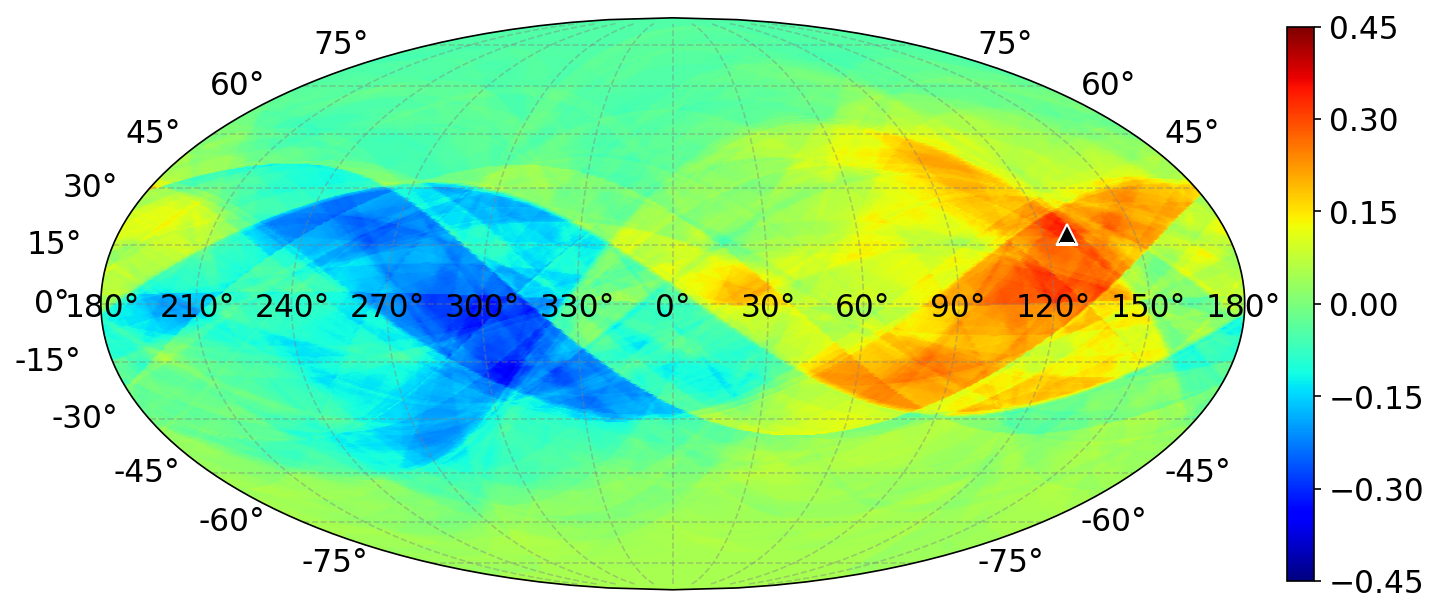}
		\caption{Excluding DES}
		\label{fig:HC_eDES}
	\end{subfigure}
	\hfill
	\begin{subfigure}[b]{0.49\textwidth}
		\centering
		\includegraphics[width=\textwidth,height=0.21\textheight,keepaspectratio]{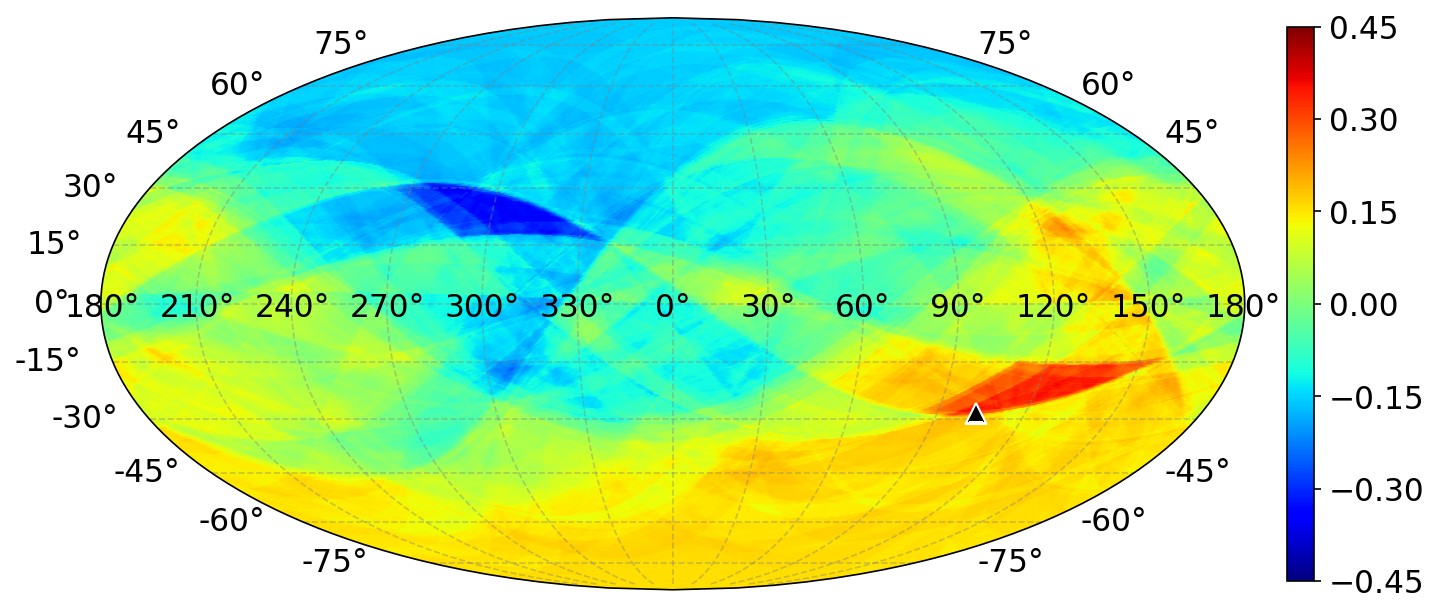}
		\caption{Excluding High-$z$}
		\label{fig:HC_eHighz}
	\end{subfigure}
	\caption{Pseudo-color maps of the anisotropy level in the Galactic coordinate system, obtained via the HC method for the full Pantheon+ sample, as well as the cases excluding the Low-$z$, PS1, SDSS, SNLS, DES, and High-$z$ subsamples, respectively. The triangle in each panel marks the direction of maximum anisotropy level.}
	\label{fig:HC1}
\end{figure*}

For the HC procedure, the \texttt{healpy}\footnote{http://healpix.sourceforge.net} Python package \citep{Gorski:2004by, Zonca:2019vzt} is employed to produce the random orientations. With the parameter $N_{\mathrm{side}}$ set to 128, this resolution corresponds to a total of $12 \times N_{\mathrm{side}}^2=196,608$ directions. Figure \ref{fig:HC1} presents the pseudo-color maps of the anisotropy level, while the numerical results are compiled in Table \ref{table:HC}.

\begin{table*}[!htbp]
	\centering
	\caption{Results of the HC method. This table presents the maximum anisotropy level ($\mathrm{AL}_{\mathrm{max}}$), its associated direction $(l, b)$ in the Galactic coordinate system, the best-fit values of $\Omega_{\mathrm{m}}$ in the ``up'' and ``down'' hemispheres, the total number of SNe Ia ($N$), and the sample sizes within each hemisphere. The numbers in parentheses denote the additionally retained Cepheid-hosting SNe Ia.}
	\label{table:HC}
	\setlength{\tabcolsep}{2.85mm}
	\renewcommand{\arraystretch}{1.3}
	\resizebox{\textwidth}{!}{%
		\begin{tabular}{lcccccccc}
			\hline
			Sample & $N$ & $N_u$ & $N_d$ & $\Omega_{\mathrm{m}, u}$ & $\Omega_{\mathrm{m}, d}$ & $\mathrm{AL}_\mathrm{max}$ & $l\,[^\circ]$ & $b\,[^\circ]$ \\
			\hline
			Full Pantheon+
            & $1701$
            & $1208$
            & $493$
            & $0.370 \pm {0.023}$
            & $0.276 \pm {0.024}$
            & $0.289 \pm {0.052}$
            & $127.97_{-7.03}^{+9.84}$
            & $17.90_{-18.19}^{+3.49}$ \\

			Excluding Low-$z$
            & $983\,(77)$
            & $691\,(26)$
            & $292\,(51)$
            & $0.402 \pm {0.033}$
            & $0.261 \pm {0.031}$
            & $0.426 \pm {0.068}$
            & $63.63_{-129.02}^{+15.12}$
            & $-18.21_{-6.41}^{+34.54}$ \\

			Excluding PS1
            & $1432$
            & $991$
            & $441$
            & $0.378 \pm {0.025}$
            & $0.282 \pm {0.025}$
            & $0.291 \pm {0.053}$
            & $128.67_{-7.73}^{+18.28}$
            & $17.90_{-18.19}^{+4.13}$ \\

			Excluding SDSS
            & $1380$
            & $898$
            & $482$
            & $0.369 \pm {0.024}$
            & $0.278 \pm {0.024}$
            & $0.280 \pm {0.053}$
            & $127.62_{-71.72}^{+15.47}$
            & $18.21_{-46.84}^{+3.81}$ \\

			Excluding SNLS
            & $1541$
            & $1190$
            & $351$
            & $0.358 \pm {0.024}$
            & $0.239 \pm {0.042}$
            & $0.399 \pm {0.080}$
            & $128.67_{-29.53}^{+14.77}$
            & $-21.06_{-7.91}^{+3.79}$ \\

			Excluding DES
            & $1498$
            & $1103$
            & $395$
            & $0.362 \pm {0.024}$
            & $0.249 \pm {0.026}$
            & $0.369 \pm {0.059}$
            & $127.97_{-14.06}^{+8.79}$
            & $17.90_{-18.79}^{+4.13}$ \\

			Excluding High-$z$
            & $1671$
            & $1193$
            & $478$
            & $0.379 \pm {0.024}$
            & $0.262 \pm {0.031}$
            & $0.367 \pm {0.061}$
            & $103.36_{-7.38}^{+41.48}$
            & $-28.29_{-1.71}^{+11.33}$ \\

			SNLS
            & $160\,(77)$
            & $107\,(33)$
            & $53\,(44)$
            & $0.554 \pm {0.132}$
            & $0.234 \pm {0.107}$
            & $0.813 \pm {0.216}$
            & $132.19_{-7.03}^{+24.61}$
            & $16.02_{-26.53}^{+6.32}$ \\

			Low-$z$+High-$z$
            & $748$
            & $349$
            & $399$
            & $0.487 \pm {0.090}$
            & $0.253 \pm {0.037}$
            & $0.634 \pm {0.132}$
            & $38.32_{-85.78}^{+14.06}$
            & $-1.79_{-19.27}^{+36.74}$ \\

			Low-$z$+SNLS+High-$z$
            & $908$
            & $603$
            & $305$
            & $0.360 \pm {0.028}$
            & $0.255 \pm {0.030}$
            & $0.341 \pm {0.067}$
            & $129.38_{-27.07}^{+26.37}$
            & $18.52_{-20.02}^{+5.77}$ \\
			\hline
		\end{tabular}%
	}
\end{table*}

\begin{figure}[!htbp]
	\begin{center}
		\includegraphics[width=7.5cm]{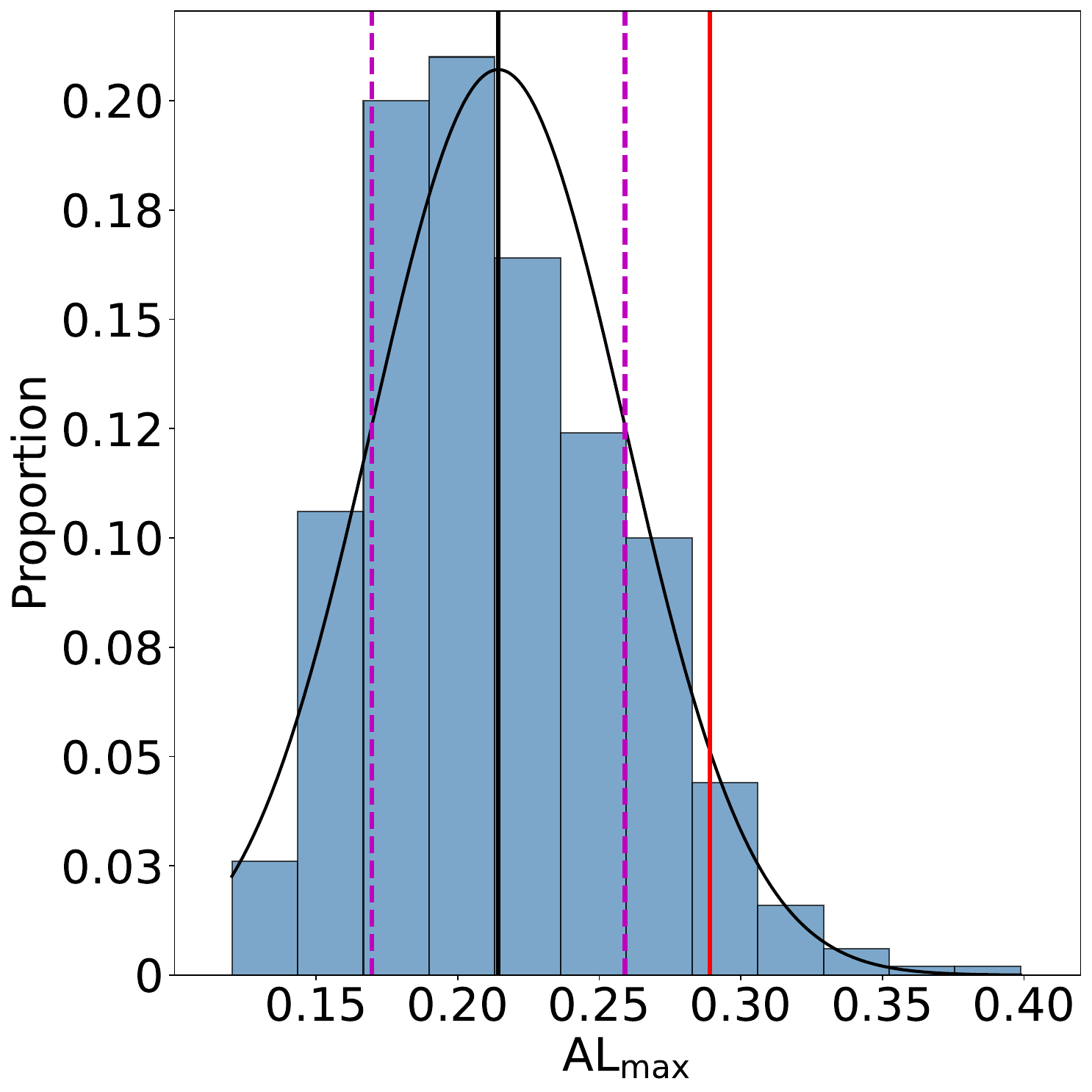}
		\caption{Distribution of the maximum anisotropy levels ($\mathrm{AL}_{\mathrm{max}}$) derived from the mock simulations (histogram), overlaid with the best-fit Gaussian function (black curve). The solid and dashed vertical lines denote the mean and standard deviation of the mock distribution, respectively, while the red vertical line marks the observed $\mathrm{AL}_{\mathrm{max}}$ value from the full Pantheon+ sample.}
		\label{fig:HC_mock}
	\end{center}
\end{figure}

We first apply the HC method to the full Pantheon+ sample. For this complete dataset, the maximum anisotropy level is found to be $\mathrm{AL}_{\mathrm{max}}=0.289 \pm 0.052$, with a preferred direction of $(l,b)=(127.97^\circ,17.90^\circ)$. We conduct an isotropic mock analysis to assess the statistical significance of the maximum anisotropy level in the full sample. By randomizing only the sky positions of the SNe Ia while keeping all other properties identical to the actual data, we generate 500 mock realizations. The maximum anisotropy levels derived from the mock simulations are plotted in Figure \ref{fig:HC_mock}. The resulting $p$-value is found to be 0.060, which corresponds to a Gaussian-equivalent one-sided significance of $1.56\sigma$. This indicates that the evidence for anisotropy in the full sample is statistically insignificant.

Next, we systematically exclude each subsample in turn and compare the corresponding results with the full-sample result. Excluding the Low-$z$ subsample leads to an increase in the maximum anisotropy level to $\mathrm{AL}_{\mathrm{max}}=0.426 \pm 0.068$, which deviates from the full-sample value by approximately $1\sigma$. Conversely, when the PS1, SDSS, SNLS, DES, or High-$z$ subsample is excluded, the obtained $\mathrm{AL}_{\mathrm{max}}$ values remain consistent with the full-sample result within the $1\sigma$ uncertainties. These findings suggest that the maximum anisotropy level is particularly sensitive to the exclusion of the Low-$z$ subsample.

The preferred direction also exhibits varying sensitivity to the exclusion of different subsamples. Specifically, excluding the Low-$z$, SNLS, or High-$z$ subsample shifts the preferred directions to $(63.63^\circ,-18.21^\circ)$, $(128.67^\circ,-21.06^\circ)$, and $(103.36^\circ,-28.29^\circ)$, respectively. The angular separations from the full-sample direction are approximately $72.82^\circ$, $38.97^\circ$, and $51.96^\circ$, respectively. These shifts are significantly larger than those observed upon excluding the PS1, SDSS, or DES subsamples. In these latter cases, the preferred directions remain closely aligned with that of the full sample, showing angular separations of only $0.67^\circ$, $0.45^\circ$, and $0^\circ$, respectively. Consequently, the PS1, SDSS, and DES subsamples exert a negligible impact on the preferred direction of the full Pantheon+ sample.

To further investigate the role of these subsamples, we fix the orientation to the preferred direction obtained from the full Pantheon+ sample, approximately $(l,b)=(127.97^\circ,17.90^\circ)$. We then calculate the anisotropy levels of the Low-$z$, SNLS, and High-$z$ subsamples along this axis. The resulting values are $\mathrm{AL}_{\mathrm{Low}\text{-}z}=-1.385$, $\mathrm{AL}_{\mathrm{SNLS}}=0.376$, and $\mathrm{AL}_{\mathrm{High}\text{-}z}=-0.737$. The negative results obtained for the Low-$z$ and High-$z$ cases reveal that these two subsamples provide opposing contributions along this preferred direction. Consequently, they act to suppress the anisotropy level, whereas the SNLS subsample provides a positive contribution.

\begin{figure}[!htbp]
	\centering
	\begin{subfigure}[b]{0.49\textwidth}
		\centering
		\includegraphics[width=\textwidth]{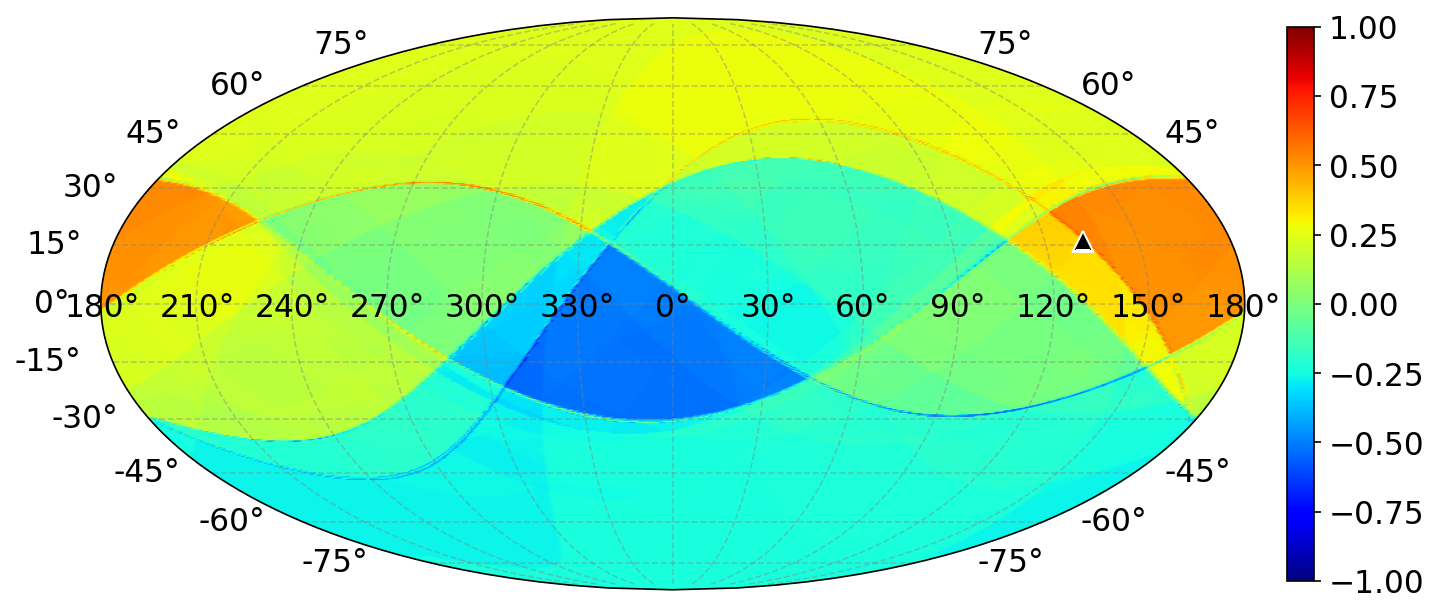}
		\caption{SNLS}
		\label{fig:HC_SNLS}
	\end{subfigure}
    \begin{subfigure}[b]{0.49\textwidth}
		\centering
		\includegraphics[width=\textwidth]{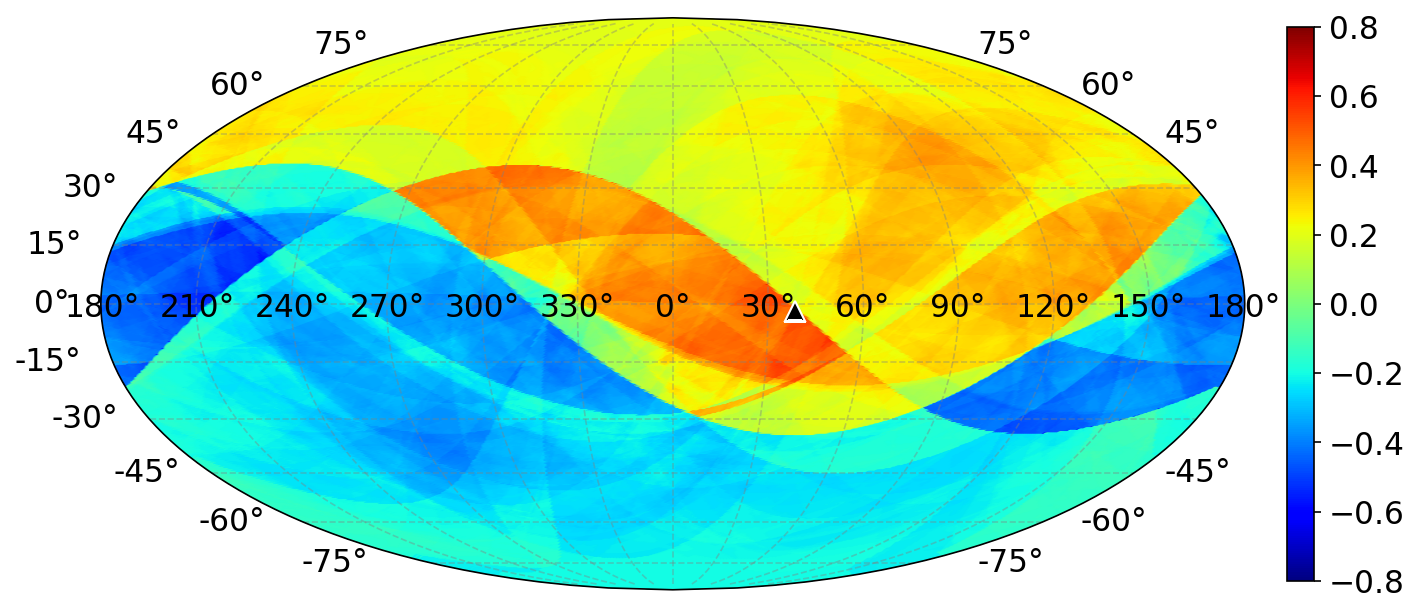}
		\caption{Low-$z$+High-$z$}
		\label{fig:Lowz_Highz}
	\end{subfigure}
    \begin{subfigure}[b]{0.49\textwidth}
		\centering
		\includegraphics[width=\textwidth]{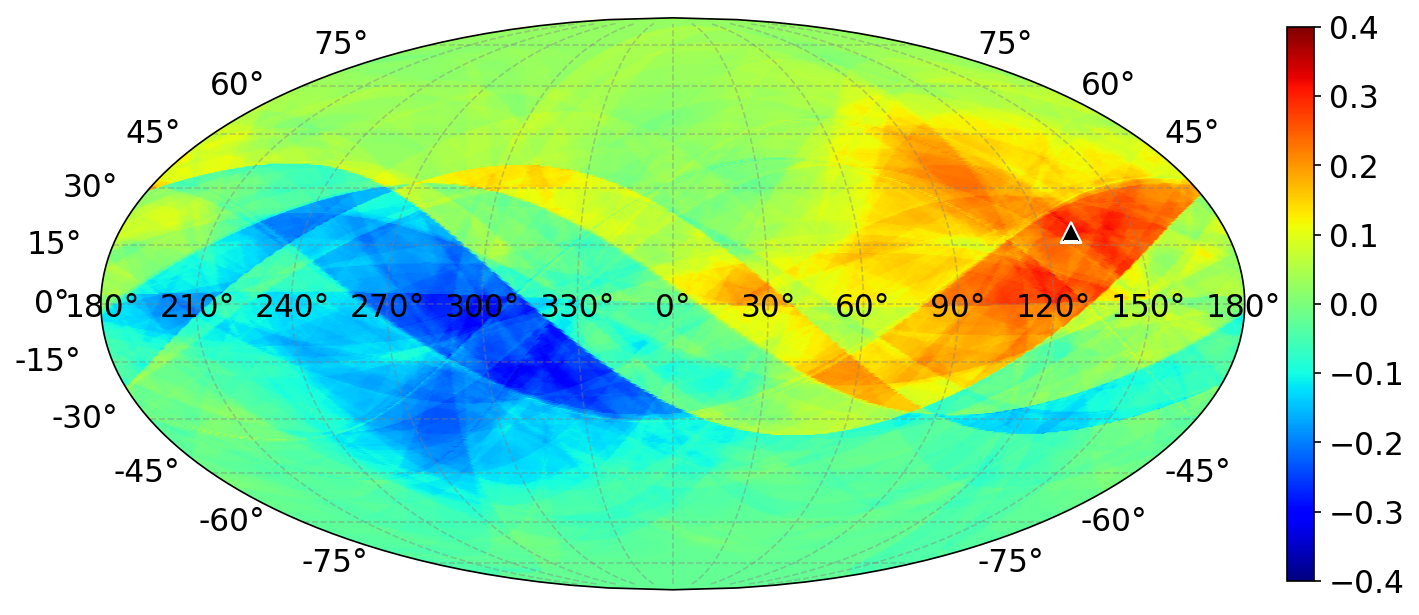}
		\caption{Low-$z$+SNLS+High-$z$}
		\label{fig:Lowz_Highz_snls}
	\end{subfigure}
	\caption{Pseudo-color maps of the anisotropy level in the Galactic coordinate system, obtained via the HC method for the SNLS subsample, the combined Low-$z$ and High-$z$ dataset, and the combined Low-$z$, SNLS, and High-$z$ dataset, respectively. The triangle in each panel marks the direction of maximum anisotropy level.}
	\label{fig:HC2}
\end{figure}

We also perform an independent analysis on the SNLS subsample using the HC method. For this subsample, we obtain $\mathrm{AL}_{\mathrm{max}}=0.813 \pm 0.216$ with a preferred direction of $(l,b)=(132.19^\circ,16.02^\circ)$. This direction is closely aligned with the full-sample preferred direction, showing an angular separation of only $4.45^\circ$. This suggests that the preferred direction of the full Pantheon+ sample is strongly influenced by the SNLS subsample. Figure \ref{fig:HC2} presents the pseudo-color map of the SNLS anisotropy level, with the numerical results compiled in Table \ref{table:HC}.

On the other hand, the combined Low-$z$ and High-$z$ subset yields $\mathrm{AL}_{\mathrm{max}}=0.634 \pm 0.132$ with a preferred direction of $(l,b)=(38.32^\circ,-1.79^\circ)$. The angular separation between this direction and the full-sample preferred direction is approximately $90.22^\circ$. This indicates that the combined Low-$z$ and High-$z$ subsamples alone cannot reproduce the full-sample preferred direction. Figure \ref{fig:HC2} presents the pseudo-color map of the anisotropy level for the combined Low-$z$ and High-$z$ subsamples, with the numerical results compiled in Table \ref{table:HC}.

Finally, when the SNLS subsample is added to the combined Low-$z$ and High-$z$ subsamples, we obtain $\mathrm{AL}_{\mathrm{max}}=0.341 \pm 0.067$ with a preferred direction of $(l,b)=(129.38^\circ,18.52^\circ)$. Figure \ref{fig:HC2} presents the pseudo-color map of the anisotropy level for the combined Low-$z$, SNLS, and High-$z$ subsamples, with the numerical results compiled in Table \ref{table:HC}. This direction is closely aligned with that of the full Pantheon+ sample, showing an angular separation of only $1.48^\circ$. Therefore, these results suggest that the preferred direction of the full Pantheon+ sample is primarily determined by the SNLS subsample. Meanwhile, the Low-$z$ and High-$z$ subsamples act to suppress the anisotropy level along this direction.

\section{Conclusion}\label{Conclusion}
In this work, we employed both the dipole fitting (DF) and hemisphere comparison (HC) methods to probe cosmic anisotropy within the Pantheon+ sample.

For the dipole fitting (DF) method, we found that the anisotropic signal within the full Pantheon+ sample is statistically weak. The associated dipole direction is oriented towards $(l,b)=(138.39^\circ,8.53^\circ)$. After we excluded the Low-$z$ subsample, we found that this direction shifts to $(l,b)=(334.02^\circ,30.63^\circ)$, representing an angular separation of $138.07^\circ$ from the original preferred direction. Conversely, when we excluded the PS1, SDSS, SNLS, DES, or High-$z$ subsamples individually, the resulting constraints remain broadly consistent with those of the full Pantheon+ sample. We further analyzed the Low-$z$ subsample independently. For this subsample, the dipole amplitude is $A_{\mathrm{D}}=0.952^{+0.454}_{-0.403}\times10^{-3}$ with a preferred direction of $(l,b)=(149.77^\circ,-12.20^\circ)$. To identify the origin of this signal, we investigated the individual surveys constituting the Low-$z$ subsample. We found that the dipole signal predominantly arises from surveys 5, 56, 63, and 150. The combined subset of these surveys yields $A_{\mathrm{D}}=1.730_{-0.715}^{+0.554}\times10^{-3}$ with a preferred direction of $(l,b)=(153.05^\circ,-1.25^\circ)$. This direction lies only $11.42^\circ$ away from that of the full Low-$z$ subsample. To evaluate the significance of these dipole signals, we performed 5000 isotropic simulations using the $\Delta\chi^2$ statistic. These tests were applied to both the Low-$z$ subsample and the combined subset of surveys 5, 56, 63, and 150. The resulting significance is $2.27\sigma$ for the Low-$z$ subsample and $2.79\sigma$ for the combined subset. Both datasets span the redshift range of $0.001 < z < 0.093$. This is consistent with previous studies \citep{Tang:2023kzs, Sorrenti:2022zat, Cowell:2022ehf, Sorrenti:2024ugq, Sah:2024csa} reporting a dipole signal of approximately $2\sigma$ at low redshifts.

For the hemisphere comparison (HC) method, the full Pantheon+ sample yields $\mathrm{AL}_{\mathrm{max}}=0.289\pm0.052$ with a preferred direction of $(l,b)=(127.97^\circ,17.90^\circ)$. This result has a significance of $1.56\sigma$, suggesting weak evidence for anisotropy in the full sample. After we excluded the Low-$z$ subsample, $\mathrm{AL}_{\mathrm{max}}$ deviates from the full-sample result by approximately $1\sigma$. Conversely, when the PS1, SDSS, SNLS, DES, or High-$z$ subsamples were excluded, the resulting values remain consistent with the full-sample value within $1\sigma$. The preferred direction shifts significantly when the Low-$z$, SNLS, or High-$z$ subsamples were excluded. These shifts correspond to angular separations from the original preferred direction of $72.82^\circ$, $38.97^\circ$, and $51.96^\circ$, respectively. In contrast, when we excluded the PS1, SDSS, or DES subsamples, the preferred direction remains largely unchanged. To clarify the origin of the full-sample preferred direction, we evaluated the anisotropy levels of the Low-$z$, SNLS, and High-$z$ subsamples along this axis. Along this direction, the Low-$z$ and High-$z$ subsamples yield negative contributions to $\mathrm{AL}$, whereas the SNLS subsample provides a positive one. The SNLS subsample alone yields a preferred direction of $(l,b)=(132.19^\circ,16.02^\circ)$, lying only $4.45^\circ$ away from the full-sample direction. Moreover, the preferred direction of the combined Low-$z$ and High-$z$ subsamples differs from the original direction by $90.22^\circ$. This separation decreases to only $1.48^\circ$ after we added the SNLS subsample. Consequently, these results suggest that the preferred direction of the full Pantheon+ sample is primarily determined by the SNLS subsample, whereas the Low-$z$ and High-$z$ subsamples act to suppress the anisotropy level along this axis.

A combination of the DF and HC analyses reveals no robust evidence for cosmic anisotropy in the full Pantheon+ sample. The DF signal is primarily associated with the Low-$z$ subsample, while the HC preferred direction is largely driven by the SNLS dataset. These subsample-dependent features favor local structures or the inhomogeneous distribution of the datasets over an intrinsic cosmic anisotropy.

\backmatter
\bmhead{Acknowledgements}
Yong Zhou is supported by the National Natural Science Foundation of China (Grant No. 12405077).

\bibliography{sn-bibliography}

\end{document}